# Transverse quantum-state characterization of programmable electron optics

S. You,[1] P. Rosi,[2] E. Rotunno,[2] A. Roncaglia,[3] L. Belsito,[3] A. H. Tavabi,[4] R. E. Dunin-Borkowski,[4] V. Grillo,[2,*] and P. M. Pelz[1,#]

[1]*Friedrich-Alexander-Universität Erlangen-Nürnberg, Erlangen, Germany*
[2]*CNR Istituto Nanoscienze, Modena, Italy*
[3]*CNR Istituto per lo Studio dei Materiali Nanostrutturati, Bologna, Italy*
[4]*Ernst Ruska-Centre for Microscopy and Spectroscopy with Electrons, Forschungszentrum Jülich, 52425, Jülich, Germany*

Programmable electron optics — electronically controlled phase plates — underpin proposals from dose-efficient phase imaging to shaped-electron X-ray sources, nearly all assuming a pure, fully coherent delivered wave whose purity has never been measured. Here we reconstruct the transverse density matrix of a microelectromechanical electrostatic spiral phase plate by mixed-state ptychography, from one four-dimensional STEM scan per state and without added hardware. The delivered beam is substantially mixed: its purity falls from $\approx 0.47$ to $\approx 0.24$ as the applied bias grows, inconsistent with a fixed lateral source-blur model, while the real-space coherence width stays near 1 nm. The same scans calibrate the device in situ, allow virtual orbital-angular-momentum sorting and, through a partial-coherence-aware transfer theory, indicate that purifying the output could improve dose efficiency roughly threefold. One acquisition thus becomes a quantum-state acceptance test for programmable electron optics, supplying the purity and coherence that emerging phase-plate and diffractive-imaging schemes assume but leave unquantified.

## I. INTRODUCTION

Electron vortex beams, which are free-electron states whose wavefunction carries a helical phase $e^{i\ell\varphi}$ and a quantized orbital angular momentum (OAM) of $\ell\hbar$ per electron, were predicted in 2007 [1] and first realized in the transmission electron microscope in 2010 [2,3]. Because the electron is charged, its vortex states differ fundamentally from their optical ancestors: the circulating probability current endows the beam with a magnetic moment proportional to $\ell$ [4,5], turning vortex beams into nanoscale probes of magnetism [3,6,7], chirality, and fundamental quantum dynamics such as free-space Landau and Aharonov–Bohm physics [4]. For the aperture-limited convergent probes of a scanning transmission electron microscopy (STEM) instrument, the vortex ring radius grows approximately linearly with $\ell$ [5], so the achievable charge, purity, and coherence of the beam directly set the spatial resolution and sensitivity of every such application [8].

The first decade of vortex-beam optics was built on static elements: graphite-flake spiral phase plates [2]; binary fork masks, reproducible but splitting the current over many diffraction orders [3,9]; nanofabricated holograms concentrating $\ell \approx 200\hbar$ [10] and later $\ell = 1000\hbar$ at up to 20% efficiency [11] into the first order; detuned aberration correctors giving hologram-free but impure $\ell = 1$ beams [12]; and phase kinoforms reaching nondiffracting Bessel modes [13]. All are fixed at fabrication — one charge per device — with efficiency–purity trade-offs, inaccessible central singularities and spurious orders as recurring limitations.

Electrostatic phase elements lift this rigidity. Building on the electrostatic Aharonov–Bohm effect of charged line dipoles [14] and the magnetic monopole-like field of a magnetized needle [15], programmable and tunable devices have emerged: einzel-lens phase-plate arrays with 4 [16] and 48 [17] independently addressable pixels, the tunable Ampère phase plate for in-focus phase contrast [18], and, most relevant here, the microelectromechanical system (MEMS) electrostatic spiral phase plate (SPP), which generates isolated electron vortex beams whose charge is continuously tunable by the bias applied across a single electrode pair up to more than $1000\hbar$ [19]. These devices place the OAM of the probe under electronic control, without sacrificial diffraction orders and with near-lossless transmission.

Programmable electron optics has, in turn, inspired a rapidly growing catalog of proposals: illumination approaching the quantum limit of dose efficiency, extracting several times more phase information per electron than an ideal Zernike plate [20,21]; probes that detect chiral phonons [22] and obey atomic-like selection rules mapping nano-optical multipole moments [23]; wavefunctions shaped to enhance coherent X-ray emission, an 18-fold gain already proposed for an existing 48-pixel plate [24,25]; stacked plates performing arbitrary conformal mappings [26,27]; and a nascent free-electron quantum optics in which shaped electrons entangle with atoms [28] and herald quantum light of designed symmetry [21].

*Contact author: vincenzo.grillo@unimore.it
#Contact author: philipp.pelz@fau.de

Characterization, however, has kept pace with neither generation nor ambition. The delivered beam has historically been verified by intensity-based signatures — doughnut diffraction patterns and defocused edge dislocations [3], through-focus evolution of the dark core [9], or knife-edge Gouy rotations giving a mean charge to $\approx 12\%$ at large $\ell$ [11] — rather than measured directly on the beam [10,13]. Even dedicated OAM diagnostics deliver only partial projections of the state. The log-polar OAM sorter is the most complete of them: it performs a rigorous projective measurement of the OAM populations $|c_\ell|^2$ at single-electron efficiency, valid for pure and mixed states alike and free of any assumption about the beam [29], and the conjugate log-radial coordinate of the same conformal mapping gives access to the radial (scaling) spectrum as well [30]. Being an intensity measurement, however, it returns the diagonal of the OAM density matrix — with tens-of-percent adjacent-channel cross-talk set by the finite azimuthal extent over which the beam is unwrapped [29] — and not the relative phases of the coefficients or the off-diagonal coherences between channels. The remaining methods project further still: spiral-phase-plate imaging filters one OAM channel per exposure and presumes a pure state [31,32]; interference-rotation magnetometry compresses the beam into a single rotation observable [7]; diffraction holography has retrieved the coherent wavefront (phase) of a hologram-generated vortex beam [33], but recovers a single pure wave and does not access its incoherent mode content or off-diagonal coherence; and off-axis holography has reconstructed the coherent needle-plane phase of the MEMS SPP itself, but only up to a few hundred quanta (extrapolated linearly beyond) and again as a single pure wave [19]. None of these methods accesses what a partially coherent source actually produces: a mixed state, whose incoherent mode content fills in the defining dark core of the vortex and becomes the limiting factor for small probes [8]. The Reviews of Modern Physics survey of the field draws the explicit conclusion: the existing diagnostics "work well for single pure OAM states", but for mixed states "a new approach is required" [5]. The application proposals sidestep the question entirely, modeling the delivered beam as a pure state transformed by an ideal, lossless phase element; the assumption is so ingrained that even quantitative dose-efficiency analyses "simply assume it to be perfect" [20]. The roadmap for quantum nanophotonics with free electrons likewise singles out measurement of the coherence-carrying off-diagonal elements of the electron density matrix as an open capability that "would open a window into previously inaccessible magnitudes" [21].

This gap is not academic: those proposals tax exactly the unmeasured properties. Elastic magnetic contrast scales linearly with $\ell$ and requires matched $\pm\ell$ pairs, with percent-level signals that survive only if the probe's dose efficiency leaves enough shot-noise budget [6] — and the initial electron magnetic circular dichroism (EMCD) demonstration of 2010 has not been reproduced since [5]. Dose-efficient phase contrast of beam-sensitive specimens with MEMS plates [18] rests on purity and coherence claims never measured. X-ray enhancement scales with the number of lattice cells inside the probe's transverse coherence area [24]; pinwheel, phase-shaped and angular-momentum-resolved spectroscopy signals are interference terms surviving only insofar as the constituent momentum components stay mutually coherent [22,23,34]; and dose-optimal illumination inherits, electron for electron, the mode impurity of the device generating it [20]. Mode powers, coherence and dose efficiency must therefore be characterized together, per charge state, on the delivered probe.

Here we show that 4D-STEM ptychography closes this gap with no additional hardware. Mixed-state ptychography [35] reconstructs an effective, finite-rank density matrix of the illuminating probe from the same scanning-diffraction data used for imaging [36,37]. We introduce a general two-stage scheme for reconstructing the delivered transverse state of a programmable phase element, demonstrated here on vortex probes; each programmed state yields an effective, finite-rank transverse mutual-coherence operator: (i) a direct, non-iterative stage fits the aberration surface — and, for a vortex, the effective phase winding — from the shifts of the bright-field disk, using a loop integral that separates the vortex circulation from scan-distortion contributions; (ii) a mixed-state ptychographic stage, initialized with the stage-one probe, refines the complex probe modes. Applying this pipeline to an electrostatic MEMS spiral phase plate across the charge series $\ell = -17 \ldots + 17$, we reconstruct the probe mode ensembles and quantify (a) mode powers as a function of OAM, (b) spatial coherence as a function of OAM in both real and Fourier space, and (c) the coherence-aware spectral signal-to-noise ratio (SSNR) and detective quantum efficiency (DQE) of each probe, from which we derive quantitative rules for dose-efficient phase-contrast imaging. The reconstructed charge tracks the programmed value linearly across the series (Fig. 4), demonstrating that a standard 4D-STEM camera suffices for transverse quantum-state characterization of programmable electron optics — converting the coherence and purity that the proposals above assume into a measured, per-charge-state specification.

## II. RESULTS

### A. Reconstructed vortex probes across the charge series

Fig. 1 shows the first four reconstructed probe modes for every charge of the series, in real space and in the aperture plane. The reconstructed dominant probe modes (mode 0) in real space display the characteristic annular intensity profiles of electron vortex beams, with the central singularity (dark core) and ring radius expanding monotonically with the programmed nominal charge $|\ell|$ (the corresponding mode intensities are shown in Supplementary Fig. S1). In the aperture plane, the reconstructed phase exhibits a clear azimuthal winding. With the scan step and diffraction-space sampling $dk$ fixed to their independently calibrated values (Sec. B.) the reconstructed charge is a clean linear function of the programmed charge across the entire series (Fig. 4). The linear voltage-to-charge relation used to program the series was transferred from a different device of the same design, and this device was not itself calibrated before the experiment. Transfer between devices cannot be taken for granted: even a single device can drift over time, for instance through oxidation of the silicon electrodes, underscoring the value of the in-situ, per-acquisition charge characterization demonstrated here.

The same reconstructions also recover the specimen. Fig. 2 shows the reconstructed object phase across the charge series, after removing a linear phase ramp, correcting the STEM scan rotation, and cropping the low-coverage reconstruction border; the retrieved specimen structure is consistent across the programmed charges, as expected for a probe-only sweep on a fixed sample.

### B. Mode powers as a function of OAM

Fig. 3 summarizes the mode-power spectra of the charge series, plotted against the fit-predicted mean OAM ($\ell = 1.68\,\ell_{\text{prog}}$; the per-scan reconstructed mean OAM is reported separately in Fig. 4 c). The dominant-mode power declines linearly with charge magnitude at $-0.014 \pm 0.002$ per $\hbar$ (fit intercept $p_0 \approx 0.58$; the measured $\ell = 0$ point is 0.64), and the purity Tr $\rho^2$ at $-0.012 \pm 0.001$ per $\hbar$ (fit intercept $\approx 0.42$; the measured $\ell = 0$ value is $\approx 0.47$). Equivalently, the von Neumann entropy $S = -\mathrm{Tr}(\rho \ln \rho)$ rises from $\approx 1.1$ nats at $\ell = 0$ (effective mode number $\exp(S) \approx 3$) to $\approx 1.6$ nats ($\approx 5$ effective modes) at the largest charge, quantifying the growing information content of the mixing (a pure probe has $S = 0$). Separately acquired round-beam reference scans reconstruct nearly pure ($p_0 = 0.90...0.98$; Supplementary Fig. S4), showing that the solver recovers a simpler round probe as nearly pure and that the vortex mixedness is not a generic reconstruction artifact — though this does not, by itself, localize its cause to the device. In a self-consistency injection–recovery test using the same solver and experimental forward model, six of eight states reproduce the injected purity within $\approx 0.04$; two which have high off-diagonal OAM coherences, converge to more-mixed solutions, indicating that for strongly off-diagonal density matrices, more and increasingly diverse measurements are needed. The monotonic purity decrease with $|\ell|$ persists across the mode-count and recovery tests; the resulting uncertainty in the absolute values is quantified in Sec. F..

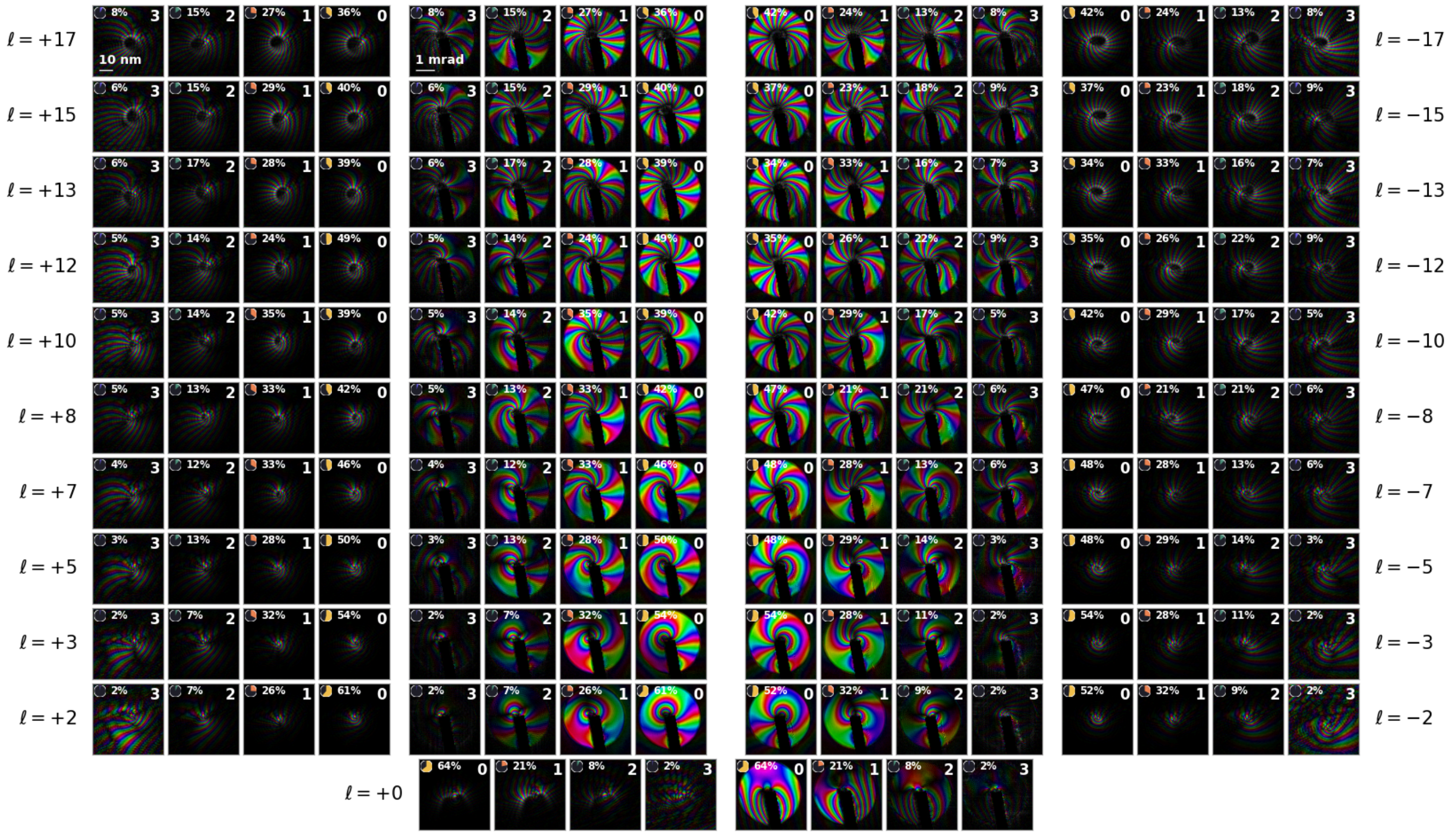

FIG. 1. First four reconstructed probe modes of the MEMS vortex probe across the charge series (one dataset per programmed charge). To keep the figure compact, opposite charges of equal magnitude share a row ($|\ell| = 17...1$), with the single $\ell = 0$ dataset centred in the final row. Each row is mirror-symmetric about its centre: from left to right it shows the real-space and aperture-plane modes of $+\ell$, then the aperture-plane and real-space modes of $-\ell$, so the two aperture-plane blocks meet at the centre and the real-space blocks sit on the outer edges beside their charge labels. The $+\ell$ (left) blocks are drawn with the mode order reversed ($3 \to 0$) so that mode 0 of both charges meets at the centre line. Modes are domain-colored (hue: phase, brightness: amplitude) and each is centred at its own intensity centre of mass. The inset pie in each panel's top-left corner shows that mode's fractional power; the mode index is printed top-right. Real-space modes are lightly cropped to their central portion, aperture modes to the aperture edge; scale bars (10 nm real-space, 1 mrad aperture) appear on the top row.

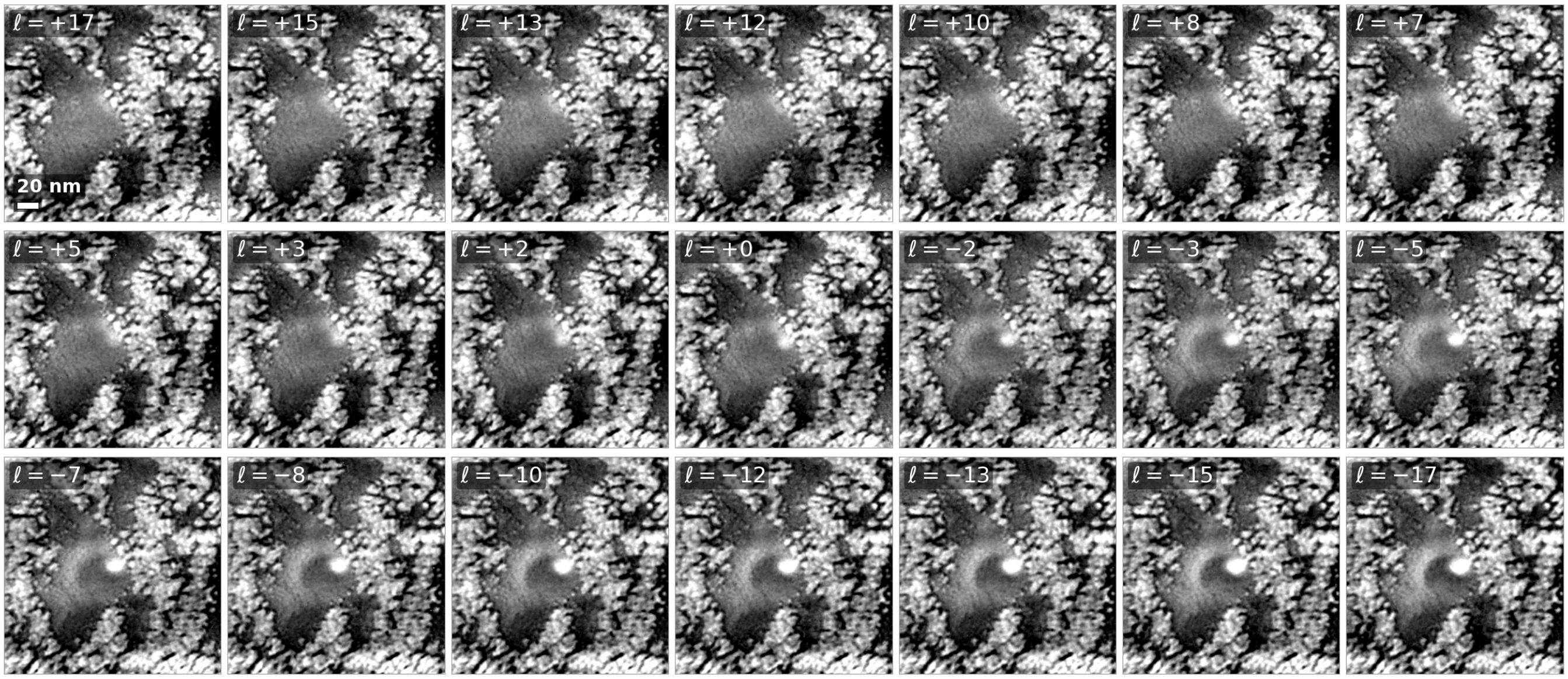

FIG. 2. Reconstructed object phase across the charge series (one dataset per programmed charge; panels are labelled by the fit-predicted mean OAM ($\ell = 1.68\,\ell_{\rm prog}$), $\ell = +17... - 17$). Each panel is the ptychographic object phase after (i) subtracting a least-squares quadratic phase surface (linear ramp plus parabola, fitted on the crop-surviving interior), (ii) de-rotating by the $1.84°$ STEM scan rotation used in the reconstruction, and (iii) cropping half a probe width ($\approx 32$ nm) from every margin to discard the low-coverage border on a common display scale ($[0.42, 0.47]$ rad, shared across all s); the reconstructions are quantitatively consistent across the charge series — the mean pairwise Fourier-ring correlation exceeds $0.5$ to a half-period of $\approx 8$ nm — as expected for a probe-only sweep on a fixed specimen. The bright spot in the center is contamination growing due to prolonged beam exposure.

### C. Virtual OAM sorting

We measure the OAM spectrum of the delivered beam computationally — a virtual analog of the hardware log-polar sorter [29,30] that additionally accesses the per-mode complex coefficients and the inter-channel coherences that an intensity measurement of the sorted plane cannot, with no dedicated column optics. Each aperture-plane mode is projected onto azimuthal harmonics about a per-dataset sorting axis, after removing the residual non-round microscope aberrations (astigmatism-dominated, median 1.3 rad at the aperture edge) that would otherwise scatter power between channels; the projection, axis refinement, and virtual aberration correction are detailed in Methods (Sec. E.).

Fig. 4 shows the resulting spectra across the charge series. The spectral ridge tracks the applied bias monotonically (panel a), and the power in the dominant OAM channel ranges from $\approx 0.20$ (at $\ell = 0$, whose round beam spreads over many OAM channels) to $\approx 0.42$. Referenced to the differential chopstick bias, the delivered ensemble-mean OAM is linear in voltage, $\langle l \rangle = (16.8 \pm 0.1\,\hbar/\mathrm{V})V + (0.6 \pm 0.1\,\hbar)$ ($R^2 = 0.999$; Fig. 4 c). Ptychography thus provides an in-situ voltage-to-mean-OAM calibration of this device–microscope–detector configuration: each volt of differential bias delivers $\approx 16.8\,\hbar$ of ensemble-mean orbital angular momentum. ($R^2$ measures linearity, not calibration accuracy; the quoted $\pm 0.1\,\hbar/\mathrm{V}$ is the regression standard error and a lower bound on the total calibration uncertainty, which also includes the reciprocal-space and scan-step calibration, voltage readback, and the sorting-axis and aberration-correction choices. A non-integer ensemble mean is not itself a topological charge. The small but nonzero intercept, $0.6 \pm 0.1\,\hbar$, most plausibly reflects a residual zero-bias offset. Expressed against the programmed integers, the delivered mean OAM scales linearly with the programmed value, slope $1.68 \pm 0.01$ ($R^2 = 0.999$), reaching $|\langle l \rangle| \approx 17$ at the extremes of the sweep. Because the gain is not an integer, the delivered charge is itself generally non-integer — and a non-integer vortex is a distinct object, not a slightly imperfect integer one. A *pure* $e^{i\ell\varphi}$ field with non-integer $\ell$ is not single-valued: it carries a radial phase discontinuity and decomposes deterministically over neighboring integer channels. Berry showed that on propagation such a beam develops a chain of alternating-charge vortices along that discontinuity rather than a single axial vortex [38], as observed for half-integer charges in light [39]. The device therefore delivers non-integer vortices by construction, and part of the measured spread over integer channels is a coherent consequence of that structure rather than evidence of mixing. Evaluated at each measured $\langle l \rangle$, this coherent decomposition accounts for $\approx 26\%$ of the measured spectral variance on average (ranging from a few percent at low charge to $\approx 65\%$ at the highest), at a purity of $\approx 0.98$. Two consequences follow. First, the mixedness inferred from the width of $P_l$ alone would be overestimated. Second, since a pure fractional vortex has $g = 1$ identically, the measured $g \lesssim 0.8$ establishes that the state is not simply a fractional vortex but a partially decohered one. Because the sorting acts on the reconstructed field rather than in the column, the residual non-round aberrations that mix adjacent channels are removed in post-processing and their effect on the spectrum is quantified directly (Sec. E.): the projection is free of the tens-of-percent channel cross-talk that the finite azimuthal unwrapping imposes on hardware log-polar sorting [29], it needs no dedicated column optics or alignment, and — as shown next — it carries the coefficient phases and inter-channel coherences that an intensity measurement in the sorted plane cannot. The price is that these quantities come from a reconstruction with a finite, chosen mode count rather than from a direct projective measurement, so the hardware sorter remains the more rigorous instrument for the populations alone; we bound that model dependence in Sec. F..

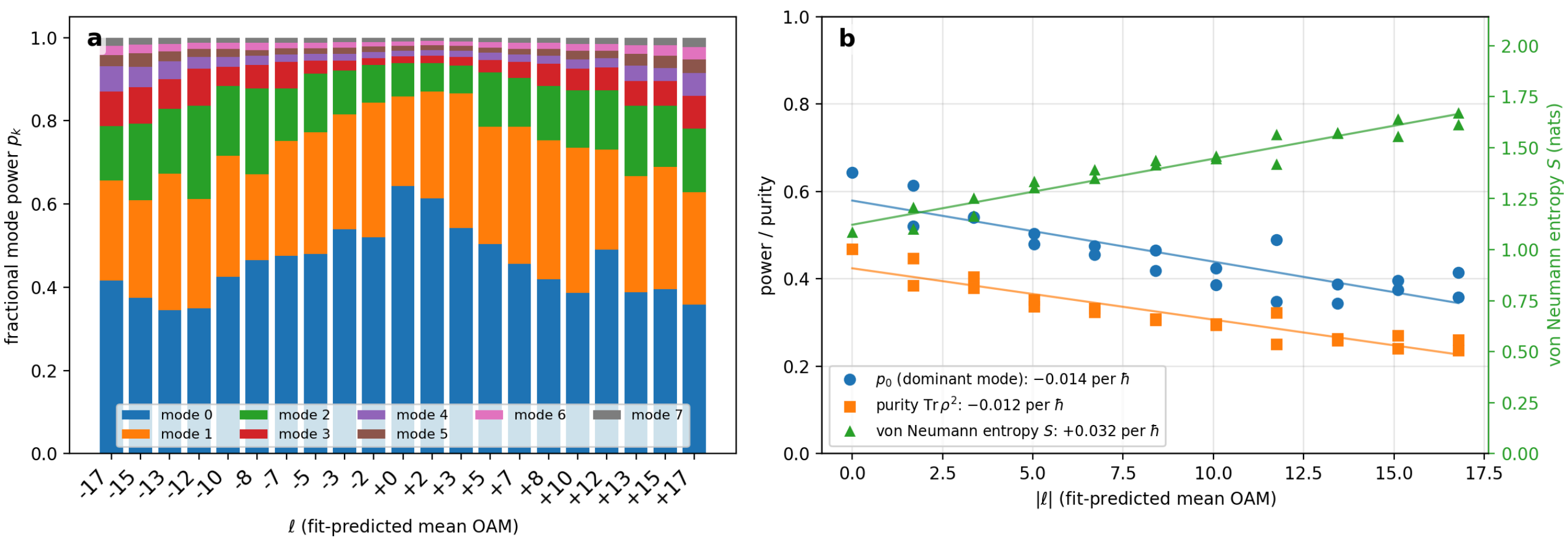

FIG. 3. Probe mode powers versus the fit-predicted mean OAM for the contiguous acquisition series (one dataset per charge; the abscissa is $\ell = 1.68 \times \ell_{\text{prog}}$, the fitted slope of Fig. 4, rather than each scan's reconstructed mean OAM). (a) Fractional mode powers $p_k$ of the eight physical probe modes (the weakest "junk" absorber mode discarded; see Sec. D.), sorted by charge. (b) Dominant-mode power $p_0$ and purity Tr $\rho^2 = \sum_k p_k^2$ (left axis) and the von Neumann entropy $S = -\sum_k p_k \ln p_k$ (right axis, green) versus charge magnitude $|\ell|$, with linear fits quantifying the trend per unit (delivered) charge: purity falls and entropy rises with charge ($S \approx 1.1 \to 1.7$ nats, effective mode number $\exp(S) \approx 3 \to 5$).

Projecting the reconstructed mutual-coherence operator onto the OAM basis (Sec. E.) resolves whether the near-integer spread is a coherent or an incoherent state. The OAM density matrix $\rho_{ll'}$ (Fig. 4 e; every dataset in Supplementary Fig. S2) carries appreciable off-diagonal weight between adjacent channels: the adjacent-OAM degree of coherence $g$ between the dominant channel and its more-populated neighbor separates the round beam sharply from every programmed vortex. Because $g$ is built from a single off-diagonal element it is not reliably determined by one reconstruction (Sec. 1.), so we quote it as an ensemble over three reconstruction seeds per state: $g = 0.78 \pm 0.03$ for the round $\ell \approx 0$ beam, falling to a *plateau* of $g = 0.30 \pm 0.03$ that is flat from $|\langle l \rangle| \approx 4$ to 17. Over the same reconstructions the diagonal quantities are stable to $\pm 0.02$ (purity) and $\pm 0.3\,\hbar$ ($\langle l \rangle$), so the sensitivity is specific to the off-diagonal. The delivered beam is therefore neither a pure OAM eigenstate nor a fully incoherent mixture, but a **partially coherent, mixed-OAM vortex** whose density matrix carries coherence between adjacent OAM components; this coherence does not by itself imply that each dominant mode is an $(\ell, \ell + 1)$ integer-OAM eigenpair, and it yields a non-integer ensemble-averaged OAM. That the coherence saturates immediately, rather than degrading in proportion to the applied bias, argues against a mechanism whose strength scales with the programmed charge and points instead to one that is switched on by the presence of winding itself.

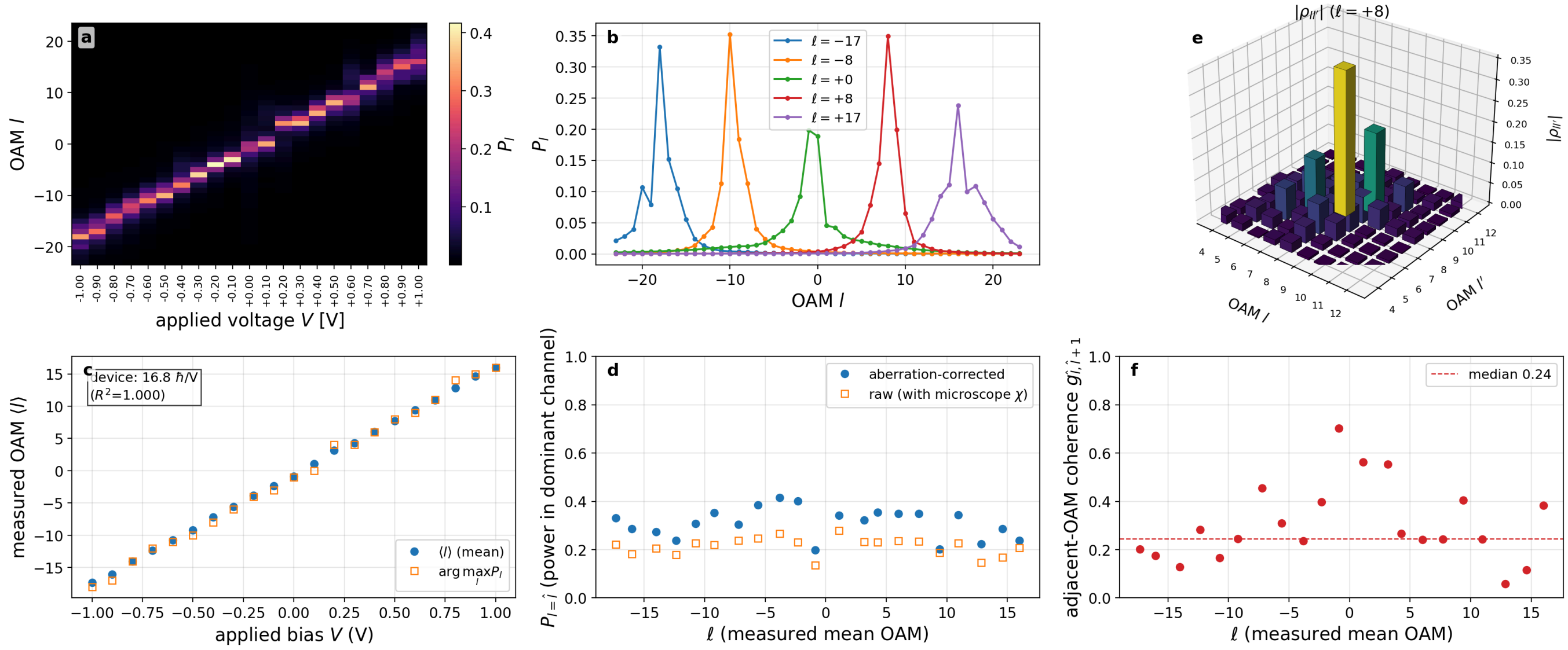

FIG. 4. Virtual OAM sorting of the reconstructed probes, obtained by projecting the reconstructed mutual-coherence operator onto the OAM basis over all coherent modes (per-dataset sorting axis and virtual aberration refinement; see Methods, Eq. (9)). (a) OAM population spectra $P_l$ across the charge series, with the applied electrode bias $V$ on the abscissa. (b) Representative spectra. (c) Device calibration: the mean $\langle l \rangle$ and dominant ($\arg\max_l P_l$) measured OAM versus the applied electrode bias $V$; the fitted slope gives the device gain $\approx 16.8\hbar/V$ (slope standard error $\pm 0.1\hbar/V$; $R^2 = 0.999$). (d) Power in the dominant channel versus the measured mean OAM. (e) OAM density matrix $|\rho_{ll'}|$ for a representative charge: the diagonal gives the populations, the off-diagonals the coherence between OAM components. (f) Adjacent-OAM degree of coherence $g$ (Eq. (10)) between the dominant channel and its neighbor versus the measured mean OAM ($\approx 0$ for an incoherent OAM mixture, $\to 1$ for a coherent superposition).

### D. Coherence as a function of OAM

Fig. 5 shows the coherence envelopes computed from the reconstructed mode stacks via Eq. (3), in real space and in the aperture plane, and their widths across the charge series. The extracted envelope widths reveal contrasting behaviors in the two domains (Fig. 5). In real space, the translationally averaged degree of coherence $|\bar{\gamma}(\Delta x)|$ is remarkably flat, staying constant at 2.92 to 3.04 pixels (corresponding to 0.96 to 1.00 nm) across all programmed charges. This flat real-space width indicates that no systematic variation of the operational sample-plane coherence width is resolved across the device states within the present reconstruction and width definition; it is a reconstruction metric, and we do not identify it with a physical source coherence length. In Fourier space, by contrast, the coherence envelope width shrinks as more winding is programmed: it falls monotonically and near-symmetrically on both branches, from $0.24\, q_{\text{probe}}$ at the smallest measured charge ($|\ell| \approx 2$) to $0.08\, q_{\text{probe}}$ at the largest ($|\ell| \approx 17$). Most of this narrowing is not a loss of angular coherence. $\bar{\gamma}$ is a *translational* average over all pairs separated by $\Delta k$ (Eq. (3)), so for an azimuthally structured state the average runs over pairs whose orientation relative to the vortex axis differs: with $P \propto A e^{i\ell\varphi}$ the integrand carries $e^{i\ell[\varphi(k)-\varphi(k+\Delta k)]}$, whose phase still varies with $k$ at fixed $\Delta k$, and the contributions partially cancel — increasingly so with $\ell$. Applying the same estimator to a *single-mode* vortex of the same charge, whose purity is exactly 1 and which by construction has no decoherence at all, reproduces almost the entire trend: the control width falls from the full aperture at $\ell = 0$ to $0.09\, q_{\text{probe}}$ at $|\ell| = 17$ (dashed curve, Fig. 5 c). Measured against it, the reconstructed states sit at $0.69 \pm 0.23$ of the control, and it is only this residual factor that carries information about coherence. The width of $\bar{\gamma}(\Delta k)$ is therefore a reconstruction diagnostic for these states rather than a measure of the long-range angular coherence that sustains a global vortex phase and its central singularity, and it should not be compared directly against the aperture diameter. The physically transparent quantities for azimuthally structured states are the equal-radius angular coherence $\Gamma(r, \varphi; r, \varphi + \Delta\varphi)$ and, equivalently, the OAM density matrix $\rho_{ll'}$ of Fig. 4 — and these disagree with $\bar{\gamma}$ in *shape*, which is the clearest evidence that the two probe different things: the adjacent-OAM coherence $g$ is a step (coherent round beam, then a charge-independent plateau) where $\bar{\gamma}(\Delta k)$ declines monotonically ($-0.90$ correlation with $|\ell|$). Because both widths are computed from the same reconstructed modes, they do not independently locate the decoherence; with the control in hand, they are consistent with a picture in which the filling-in of the vortex core at higher charges reflects aperture-plane mode mixing rather than a change in the sample-plane coherence width (which stays $\approx 1$ nm), transferring power into non-vortex modes that populate the central singularity.

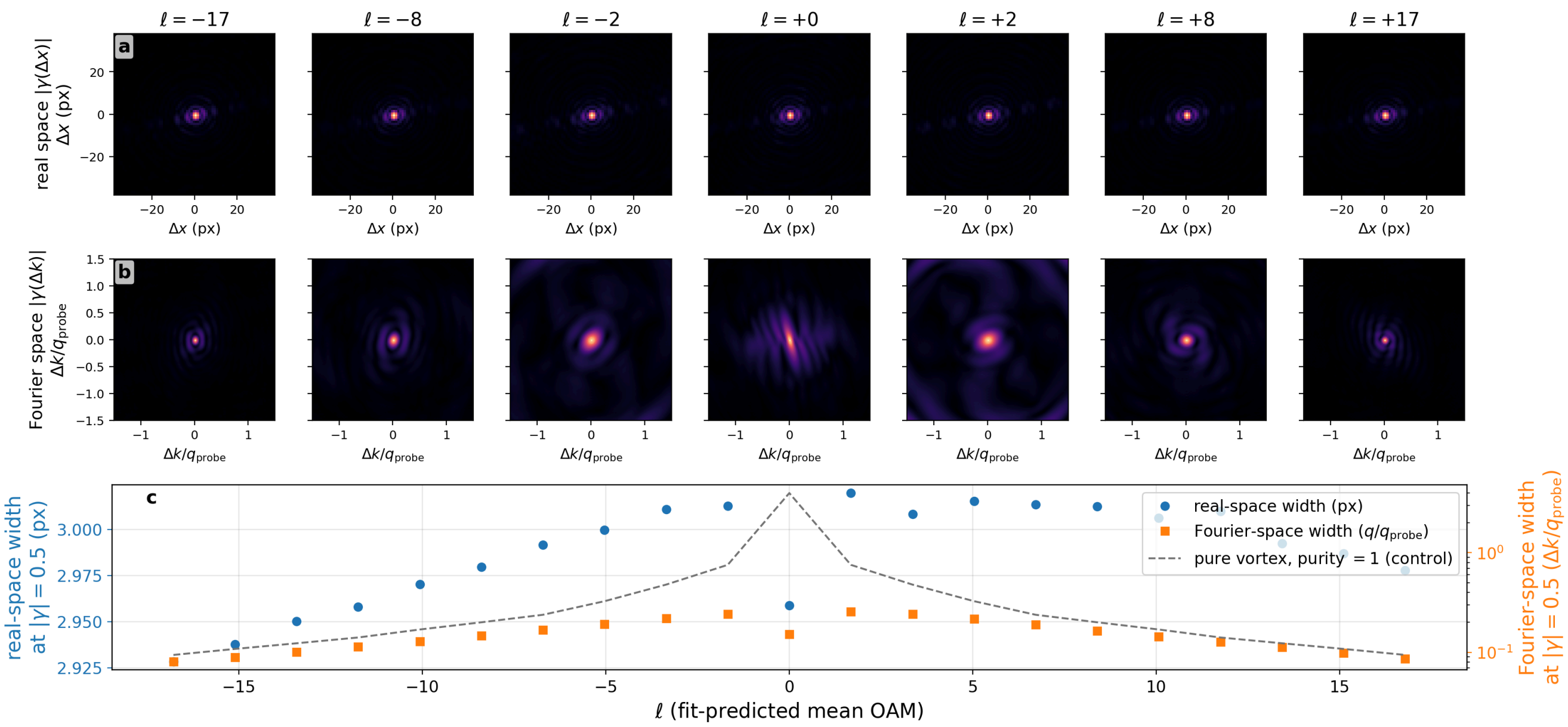

FIG. 5. Coherence envelopes of the vortex probes, following the estimation of Ref. [35] (Eq. (3)). (a) Translationally averaged degree of coherence $|\bar{\gamma}(\Delta x)|$ in the sample plane (top row) and (b) $|\bar{\gamma}(\Delta k)|$ in the aperture plane (middle row), for representative charges (columns labeled by the fit-predicted mean OAM $\ell$); (c) azimuthally averaged envelope width at $|\bar{\gamma}| = 0.5$ versus the fit-predicted mean OAM in both domains.

### E. SSNR of vortex probes compared to other available probes

Fig. 6 reports the dose-normalized direct-ptychography SSNR of the reconstructed probes across the charge series, evaluated with the partial-coherence-aware Eq. (6). The transfer separates into three regimes. First, the coherent vortex mode transfers strongly at low spatial frequencies, where the aberration-free probe is strongly damped: the vortex phase prevents the two sideband terms of $\Gamma$ from cancelling as $\boldsymbol{q} \to 0$, filling the low-$q$ band that any sufficiently structured pure aperture also fills — a defocus of twice the depth of focus ($\Delta f = 2\lambda/\alpha^2 \approx 2.2\,\mu m$) and STEM-holography illumination [40,41] (two mutually coherent probes separated by $d \approx 10$ nm, first interference passband at $q \approx 0.1\,q_{\text{probe}}$) exploit the same mechanism. Integrated over the whole band, however, this low-$q$ gain does not make the vortex the most dose-efficient pure probe: on the band-integrated $\int$ DQE $\mathrm{d}^2q$ the best coherent vortex mode reaches only $\approx 1.09\times$ the ideal round aperture and falls below the defocused aperture ($\approx 1.2\times$ the ideal), because the vortex sacrifices mid-band transfer.

Second, the measured mode mixing reduces the delivered efficiency substantially: the generalized coherence envelope of the vortex probes is $D_{\text{coh}} \approx 0.4...0.8$ across the passband, and the delivered mixed vortex reaches only $\approx 0.24...0.50$ of the ideal round aperture's dose efficiency — a factor $\approx 2.7$ below its own coherent mode. The low-$q$ transfer advantage survives partial coherence ($q \lesssim 0.5\,q_{\text{probe}}$); the integrated advantage does not.

Third, as an internal consistency check, the dedicated round-beam reference scans ($p_0 \approx 0.9$, $D_{\text{coh}} \gtrsim 0.95$; Supplementary Fig. S4) sit at or slightly above the ideal round-aperture value (not shown in Fig. 6).

## III. DISCUSSION

The measured trends constrain the factors that limit the purity of the delivered beams. Source-size blur, identified by both major reviews as the limiting factor for vortex probes [5,8], predicts improving relative coherence with increasing charge, because the vortex ring radius grows linearly with $\ell$ [5] while the effective source stays fixed. We observe the opposite: the dominant-mode power and the purity Tr $\rho^2$ decay monotonically with $|\ell|$ (Fig. 3), a trend inconsistent with a simple fixed lateral source-blur model. Other charge-dependent, source-related channels remain open — chromatic OAM dispersion from the energy-dependent electrostatic phase, angular-source averaging through the nonuniform phase element, source–device clipping and coupling, or scan instability — which the present single configuration cannot separate from a device origin. Among device-side mechanisms, bias noise and electrode charging could translate into stochastic fluctuations of the imprinted winding that incoherently mix adjacent OAM channels. ,

fixed, charge-independent absorbing mask, it can in principle affect the normalized purity of a mixed input, but being charge-independent it cannot produce the observed trend.) This candidate can be tested quantitatively. Modelling the fluctuating charge as a stochastic phase screen built from the two electrode potentials, with the electrode geometry measured rather than fitted and only the two fluctuation amplitudes free (Sec. 3.), reproduces the mixing *magnitude* — mean purity 0.26 against 0.25 measured, at 1 to 3.5 radians rms — but not its *structure*. Fitted to the populations alone, so that the coherence is a prediction, it over-predicts $g$ on the high-charge plateau by $\approx 1.3\times$ and fails qualitatively at $\ell \approx 0$, decohering the round beam to $g \approx 0.4$ where the measurement finds $0.78 \pm 0.03$; nor does the fitted amplitude grow with applied bias ($R^2 = 0.03$), as a bias-driven mechanism requires. Charging thus accounts for how much mixing there is but not for its character: it cannot hold the round beam coherent while decohering every vortex.

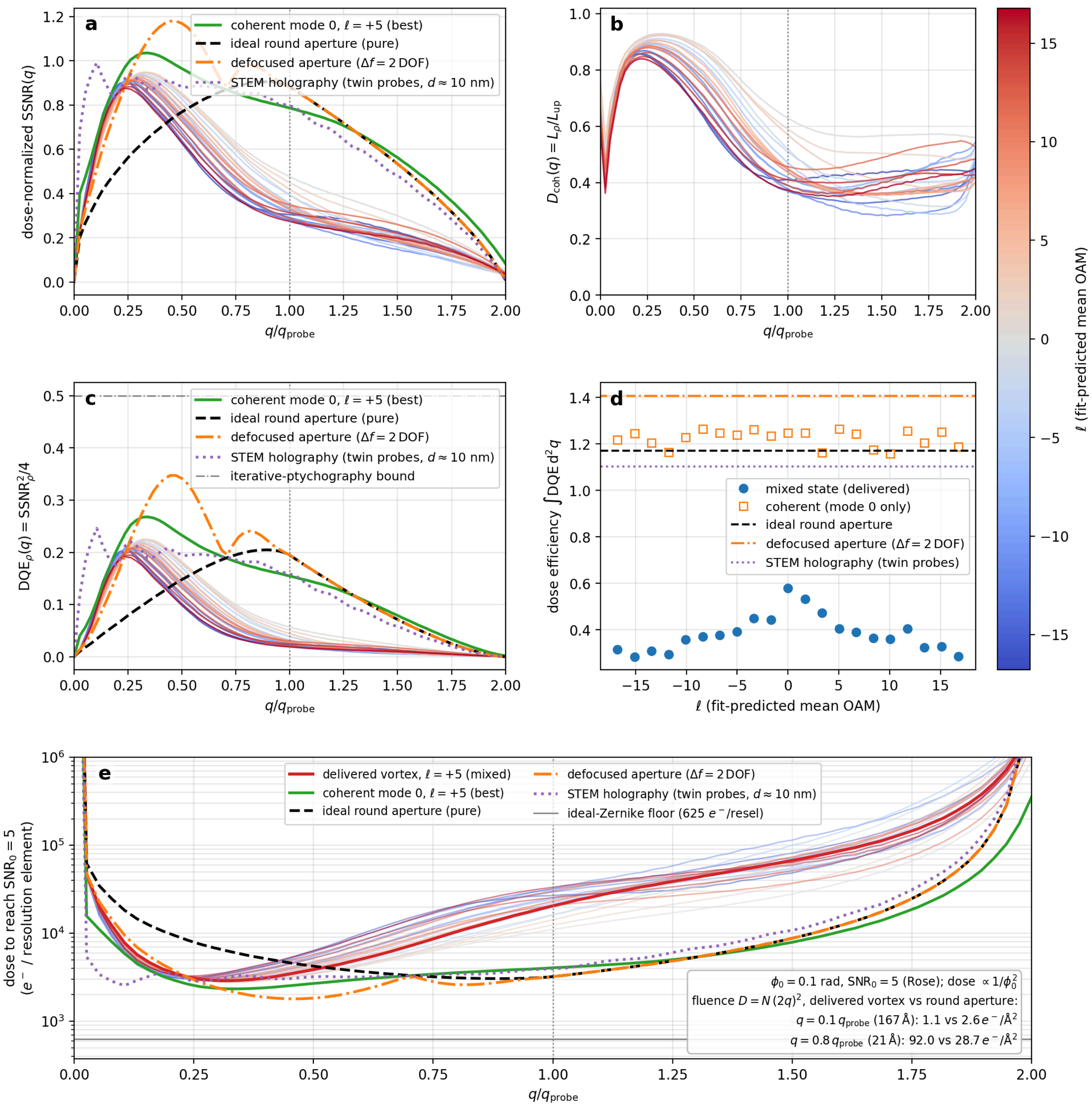

FIG. 6. Partial-coherence-aware dose-efficiency metrics for weak-phase direct ptychography for the reconstructed vortex probes (Eq. (5), Eq. (6)) [42–44]. (a) Radially averaged mixed-state $\mathrm{SSNR}_\rho(q)$ from the reconstructed physical modes, colored by the fit-predicted mean OAM; dashed curve : pure aberration-free aperture; dash-dotted orange curve: round aperture defocused by 2x DOF; dotted purple curve: STEM-holography; green curve: best fully coherent vortex mode measured with the device. (b) Generalized weak-phase coherence envelope $D_{\mathrm{coh}}(q)$ per charge. (c) DQE against the ideal Zernike reference; green curve: best fully coherent vortex mode found across the measured charge series; dash-dotted line: converged iterative-ptychography bound $\mathrm{DQE} = 1/2$ [42]. (d) Band-integrated dose efficiency $\int \mathrm{DQE}\ \mathrm{d}^2q$; dashed line: ideal round aperture, and the fractions quoted in the text are relative to it: mixed state versus the coherent dominant-mode-only value, the ideal round aperture, and the structured-probe references. (e) Dose-to-error metric: electron dose per resolution element required to reach a target phase signal-to-noise ratio $\mathrm{SNR}_0 = 5$ for a weak-phase feature of $\varphi_0 = 0.1$ rad.

Independently of these dynamic effects, the doped, conducting silicon of the electrodes bounding the beam is a candidate intrinsic source of decoherence in its own right [45,46], an effect also raised as a design concern for the $2 \times 2$ programmable phase plate [16] and applicable to the chopstick electrodes that frame the MEMS aperture here. Because doped silicon is two to three orders of magnitude less conductive than the metals for which this thermal-field decoherence was derived, however, it is expected to be a comparatively minor channel for this device. In [16], fringe visibility and low-loss EELS spectra were measured to indirectly infer decoherence. Thermal (Johnson–Nyquist) currents in any conductor close to the beam radiate a fluctuating magnetic field whose transverse component deflects the electron stochastically; averaged over the acquisition, such deflections manifest as an incoherent mixture of laterally shifted probe copies [45,46] — the laterally-shifted-mode (source-blur) channel to which $D_{\mathrm{coh}}$ reduces in the special case of incoherent laterally shifted copies (Eq. (6)); $D_{\mathrm{coh}}$ does not, in general, isolate this channel from an arbitrary recovered mixture. Because this thermal noise is set by temperature rather than by the programmed bias, it is charge-independent: it can contribute only a baseline mixedness, not the observed $|\ell|$-scaling, which must arise from a bias- or charge-dependent process. Any such baseline, scaling as $\sqrt{T}$, is in principle suppressible by cooling the device. Nor is this channel separable from electrostatic charging by the present measurement: replacing uniform charging with an independent fluctuating current in each electrode volume element leaves the recovered operator essentially unchanged, because the phase screen the electrodes can imprint has an effective rank of only $\approx 2$ however many independent sources are assumed (Sec. 3.). A detector-side contribution with the same $\ell$-scaling may also be present, and the reconstructed mode ensemble represents the combined effect of source, specimen, and detection, and a detector point-spread acting on the diffraction patterns is mathematically absorbed into probe-mode mixing [35]. Crucially, the phase information ptychography exploits is encoded indirectly in the scan-dependent intensities. Because the azimuthal phase gradient of a charge-$\ell$ vortex grows with $\ell$, the scan-dependent interference between the specimen-scattered and unscattered components can carry progressively finer structure with charge, and finite pixel integration, detector MTF, and undersampling can then remove information and bias the recovered mode rank at high $|\ell|$. This is a limitation of the measurement and reconstruction, not detector partial coherence and not physical decoherence of the delivered beam. We minimized source contributions by using the largest available spot size of our cold FEG instrument and a $50\,\mu m$ condenser aperture, and optimized the detector point-spread by $2 \times 2$ binning, which reduces but does not eliminate it (Sec. A.). Because the same probe-and-detector system characterized here is also the one at work when imaging is the goal, this finite sampling of the fine, charge-dependent diffraction is a genuine constraint for coherent diffractive vortex-probe imaging — but it limits the complete measurement system rather than the purity of the delivered beam. The virtual OAM sorter shows that a residual channel spread survives virtual correction of the astigmatism-dominated, microscope-side non-round aberrations, which is consistent with, but does not by itself establish, a device contribution.

The partial-coherence-aware SSNR (Fig. 6) links these results to dose-efficient structured illumination and diffractive imaging by quantifying how the beam properties determine dose efficiency and, when inverted per electron (Eq. (7)), the dose required to reach a target signal-to-noise ratio at each spatial frequency (Fig. 6 e). The vortex passband complements the aberration-free round aperture at low spatial frequencies but sacrifices mid-band transfer. Defocus, STEM holography, and a smaller round aperture can provide similar generic low-frequency phase contrast, so the case for vortex illumination rests on OAM-specific contrast mechanisms and on delivering the structure coherently [18,40,41]. Across the full band, the coherent vortex offers only a marginal advantage over the round aperture and falls below defocus; the recovered mixedness removes that advantage. Purifying the delivered beam would therefore yield the largest improvement in the lossless-purification model. For magnetic applications, the reconstructed powers, coherence, and OAM spectra allow matched $\pm\ell$ probe pairs to be characterized rather than assumed [5,6].

The same contingency extends beyond vortex probes to the broader program of dose-efficient diffractive and phase-plate imaging. Phase plates have just crossed from demonstration to impact — the laser phase plate now measurably improves single-particle cryo-EM structure determination of small proteins [47] at the low spatial frequencies that structured illumination also targets, and its crossed-beam successor approximates a perfect Zernike plate [48] — while a parallel line of work replaces in-focus phase contrast with diffraction under structured or random-phase illumination, solved as an inverse problem: random-phase-illuminated diffraction is estimated up to five times more dose-efficient than an ideal Zernike plate [20], and a delocalized random-aberration probe near-maximizes the quantum Fisher information for retrieving the specimen potential [49] — the regime in which low-dose cryo-electron ptychography of beam-sensitive proteins is now pursued [50]. These bounds are all derived for a fully coherent, pure incident state (and, for random illumination, an explicitly "perfect" modulator), with partial coherence bracketed out [20,49], yet coherence enters them directly: it lowers the quantum Fisher information at exactly those low spatial frequencies [49] and multiplies the DQE through the same $D_{\mathrm{coh}}^2$ envelope quantified above [43]. The promised gains are therefore only as trustworthy as the purity of the element that shapes the beam. The shortfall is exactly the mode impurity of Eq. (6), so a dose-efficiency claim for any such scheme should be accompanied by a quantum-state characterization of its shaping element, which is what the present method supplies.

The reconstructed quantities replace assumed inputs to the application proposals of the Introduction. The clearest example is the shaped-electron X-ray source: the bremsstrahlung enhancement of Ref. [24] scales linearly with the number of lattice unit cells inside the electron wavefunction's transverse coherent cross section. The real-space coherence envelope measured here corresponds to a coherence patch covering roughly ten to twenty unit cells of a typical two-dimensional crystal. As an illustrative transverse-coherence scaling this is consistent with enhancements of order ten for beams like those delivered here — comparable to the 18-fold value proposed in Ref. [24] for the existing 48-element phase plate — while the $10^3$ regime would require transverse coherence lengths approaching ten nanometers. This is a coherence-area scaling only: the actual enhancement also depends on the lattice geometry, interaction volume, phase matching, and longitudinal coherence, and the coherence area itself depends on the illumination optics through the source demagnification, so the figure is illustrative rather than a verdict on the proposal. The point is that its key coherence input can now be measured for any configuration instead of assumed. The same accounting applies to interference-based schemes: pinwheel probes for chiral-phonon detection [22] and phase-shaped spectroscopies with atomic-like selection rules [23] rely on interference between tailored momentum components whose contrast is bounded by the mutual coherence measured here (dominant-mode power 0.34...0.64 across the series, coherence envelope $D_{\mathrm{coh}} \approx$ 0.4...0.8). In each case the per-state density matrix converts an assumed-perfect input into an engineering budget.

## IV. CONCLUSIONS

We have shown that an effective transverse density matrix of a tunable electron-optical element — its mode powers, spatial coherence, and dose efficiency — can be reconstructed, per programmed state, from a single standard 4D-STEM scan, with no dedicated column hardware (the device-level conclusions draw on the full charge series and control scans). Applied to a MEMS electrostatic spiral phase plate across a large operating range, this replaces quantities the field has had to assume with measured, per-state specifications. The device delivers a mean OAM that scales linearly with the applied bias — but as a substantially mixed, partially coherent state whose purity $\mathrm{Tr}\,\rho^2$ falls — and von Neumann entropy $S$ rises — with charge, a change that these single-configuration data bound only as an upper limit on any device-generated contribution, without identifying its physical origin, since source, specimen, and detector contributions are not separately measured; the real-space coherence width stays near 1 nm. The accompanying partial-coherence-aware transfer theory converts these properties into frequency-resolved dose-to-SNR requirements and shows that the coherent vortex mode's dose-efficiency advantage over an ideal aperture is confined to low spatial frequencies and is largely erased by the measured mode mixing. More broadly, the approach is a quantum-state acceptance test for programmable electron optics: any shaped-beam device [16,17] can in principle be operationally characterized — its mode powers, coherence, OAM spectra, and dose efficiency measured before its wavefunction is trusted in an experiment — though we demonstrate this here for a single MEMS vortex device. This is the prerequisite the vortex-beam literature has repeatedly called for [5,8], and it is what the applications now driving the field require: from dose-budgeted vortex magnetometry and matched $\pm\ell$ EMCD pairs to the random-phase diffractive-imaging schemes whose promised dose gains are contingent on a purity that, as we show, a real device need not deliver. Because the reconstruction is entirely computational, the same acquisition supports closed-loop operation, feeding the measured mode spectrum back into the bias programming, pointing toward adaptive, self-calibrating electron optics whose delivered quantum state is programmed, measured, and corrected in situ.

The same measurement also indicates where the delivered state can be improved. The electrodes are not a small perturbation on this beam: their projected span is 0.27 to $0.35\,q_{\mathrm{probe}}$, so they obstruct and perturb a substantial fraction of the illuminated aperture. Within the stochastic-screen model, the mixing this produces scales with that fraction — widening the modelled electrode span from the measured value to four times it lowers the purity from $\approx 0.26$ to $\approx 0.13$ — so the ratio of electrode width to illuminated diameter is a design variable, not a fixed property of the concept. Illuminating a larger area of the device plane, or a device with narrower electrodes and a wider gap, should therefore deliver a purer vortex at the same programmed charge. We state this as a scaling trend rather than a predicted purity, since the span was varied at otherwise fixed geometry; and we expect it to act on the purity rather than on the coherence, whose step at the onset of winding is not reproduced by any mechanism we have tested. That such a prediction can be made at all is the point of the measurement: with the transverse density matrix in hand, device design becomes an optimization against a measured quantum state rather than against an assumed one.

## V. METHODS

### A. MEMS spiral phase plate and 4D-STEM acquisition

Experiments were performed on a Thermo Fisher Spectra 30–200 (S)TEM operated at 200 keV. The MEMS electrostatic spiral phase plate [19] was mounted in an in-situ current–voltage biasing holder inserted in the condenser aperture plane, so that the programmed bias imprints the vortex winding directly onto the C3 aperture. The semi-convergence angle was $\alpha = 1.5$ mrad. For each applied chopstick bias in the sweep $V = -1.0... + 1.0$ V a $256 \times 256$ position scan was recorded on a Dectris ARINA hybrid-pixel detector ($192 \times 192$ pixels) at a frame rate of 3 kHz (333.3 $\mu s$ dwell time). Diffraction patterns were binned $2 \times 2$ to optimize the effective detector point-spread. Binning does not eliminate it: the single-electron charge-cluster extent (about $2 \times 2$ native pixels at 200 keV [51]) sets only a scale for the point-spread, and does not by itself determine the binned MTF, charge sharing across bin boundaries, or the resulting reconstruction bias. The full acquisition comprises 37 four-dimensional datasets; the contiguous charge series of 21 scans used in the main-text figures samples each programmed charge once. The remainder are earlier repeat acquisitions at $\ell \leq 0$, round-beam reference scans ($\ell = 0$), and two calibration scans; a complete dataset manifest (stem, programmed charge, and role) accompanies the deposited data. The beam current, obtained by summing the detector counts per frame, was $\approx 110$ pA ($\approx 2.3 \times 10^5$ electrons per probe position, i.e. $\approx 1.6 \times 10^4 \, e^-/\text{Å}^2$ at the 3.8 Å scan step). The specimen was gold nanoparticles on an amorphous carbon support film — a standard microscope-alignment sample — whose radiation-hard, stable structure is consistent with the thin-specimen assumption of both reconstruction stages and with the probe-only nature of the charge sweep. The raw datasets were descan-corrected (center-of-mass plane fit per scan), and the azimuthal sector shadowed by the device's support needle was masked in the diffraction data.

The MINEON (MINiaturised Electron Optics for Nanobeams) phase plate was fabricated in silicon-on-insulator with trench isolation. Two parallel, doped-silicon "chopstick" electrodes, separated by $\approx 1\,\mu m$, protrude into a central aperture of radius $50\,\mu m$. Surrounding ring electrodes connected by a resistive divider establish a smooth azimuthal boundary potential, suppressing parasitic diffraction orders. Through the electrostatic Aharonov–Bohm effect, a differential bias $V$ imprints the spiral phase $e^{i\ell\varphi}$ with continuously tunable winding $\ell$. We swept $V \approx -1.0... + 1.0$ V; the device had not been calibrated beforehand.

### B. Stage 1: direct probe estimation from bright-field shifts

The first stage estimates the complex aperture function $A(\boldsymbol{k})e^{i\chi(\boldsymbol{k})}$ directly, without iterative phase retrieval, from the parallax shifts of the bright-field disk. Every detector pixel $\boldsymbol{k}$ inside the bright field (BF) disk forms a virtual image of the specimen, and for a thin specimen these images are mutually displaced by the local gradient of the aberration surface, $\boldsymbol{s}(\boldsymbol{k}) = (2\pi)^{-1}\nabla_{\boldsymbol{k}}\chi(\boldsymbol{k})$ (rotated by the scan–detector rotation) [52]. For a vortex probe, these shifts are large and strongly non-uniform, as the winding contributes a divergence-free circulation $\propto \ell/\,|\boldsymbol{k}|$, meaning that instead of registering each virtual image to a single reference, we measure shifts pairwise by cross-correlation between many detector-pixel pairs and synchronize them globally by solving the least-squares problem on the shift graph (a rank-deficient-safe graph-Laplacian solve, implementation inspired by the quantem package), iterated over a coarse-to-fine binning schedule.

The effective phase winding is then read from closed-loop integrals of the measured shift field at several radii $R$. Applying Green's theorem, the circulation separates into an $R$-independent part and a solid-body term collecting scan drift and residual rotation,

$$\frac{1}{2\pi}\oint_{|\boldsymbol{k}|\,=R} \boldsymbol{s}\cdot \mathrm{d}\boldsymbol{l} = \ell + c\,R^2, \tag{1}$$

so the intercept of a linear fit in $R^2$ yields the winding $\ell$ (loop segments crossing the masked needle sector are interpolated along the circle so every loop stays closed). Because the support needle excludes an azimuthal sector and hides the branch discontinuity, this fitted $\ell$ is an effective winding. Because the circulation is proportional to the product of the reciprocal- and real-space sampling, Eq. (1) ties the fitted winding to the diffraction-space sampling $\mathrm{d}k$. Finally, the aberration surface is fitted as $\chi(\boldsymbol{k}) = \ell\,\varphi_{\text{branch}}(\boldsymbol{k}) + \chi_{\text{smooth}}(\boldsymbol{k})$: the topological ramp is carried by an analytic azimuth term whose $2\pi$ branch cut is placed inside the needle wedge (a smooth basis cannot carry the winding through the unmeasured sector), and $\chi_{\text{smooth}}$ is a free-form cubic B-spline field fitted to the measured shifts by gradient descent through the forward model $\boldsymbol{s} = (2\pi)^{-1}\nabla_{\boldsymbol{k}}\chi$, excluding the needle sector. The aperture amplitude $|A(\boldsymbol{k})|$ is taken from the mean diffraction pattern. The resulting complex aperture seeds the dominant mode of the second stage at 96% of the probe power.

### C. Stage 2: mixed-state ptychographic refinement

The second stage refines the effective, finite-rank probe density matrix by mixed-state ptychography [35]: the recorded intensities are modeled as the incoherent sum over probe modes, $I(\boldsymbol{k}, \boldsymbol{R}) = \sum_m |\mathcal{F}[P_m(\boldsymbol{r} - \boldsymbol{R})\,O(\boldsymbol{r})]|^2$, with nine complex probe modes (the weakest, a free "junk"

mode that absorbs detector and edge artifacts, is discarded in analysis, leaving eight physical modes; Sec. D., and shown for the full series in Supplementary Fig. S3), a single object transmission function, and sub-pixel scan-position corrections, all optimized jointly through a differentiable forward model in the spirit of automatic-differentiation ptychography frameworks [53–55]. The data term is an amplitude mean-squared error on the $2\alpha$ Fourier-padded detector; mode 0 is seeded with the stage-one complex aperture at 96% of the probe power, the remaining modes with weak orthogonal fields, orthogonalized each pass. 30 epochs suffice for convergence at batch size 32 on one GPU.

### D. Characterization metrics

The reconstructed probe ensemble (the eight physical modes, after discarding the weakest "junk" mode) defines the density matrix $\rho = \sum_k |P_k\rangle\langle P_k|$, from which all characterization metrics follow. After the junk mode is discarded, the retained modes are renormalized to unit total power ($\rho \to \rho/\operatorname{Tr}\rho$; equivalently the mode powers $p_k = \|P_k\|^2/\sum_j \|P_j\|^2$ sum to one), so that $\operatorname{Tr}\rho = 1$ and the purity $\operatorname{Tr}\rho^2 = \sum_k p_k^2$ is well defined and summarizes the degree of mode mixing per charge state.

As a complementary basis-independent measure, we also report the von Neumann entropy $S = -\mathrm{Tr}(\rho \ln \rho) = -\sum_k p_k \ln p_k$ (in nats; the $p_k$ are the eigenvalues of $\rho$ for the orthogonalized modes) and the effective mode number $\exp(S)$; a pure probe has $S = 0$. Since $\sum_k p_k^2$ presumes orthogonal modes, we cross-check it against the gauge-invariant purity $\mathrm{Tr}(G^2)/(\operatorname{Tr} G)^2$ built from the mode Gram matrix $G_{jk} = \langle P_j | P_k \rangle$; the two agree to $\lesssim 10^{-3}$ across the series (maximum normalized inter-mode overlap $\lesssim 10^{-2}$), so the reconstructed modes are effectively orthogonal, and the purity is well defined. Following the coherence-envelope estimation of Ref. [35] (Supplementary Information), the full spatial coherence is quantified by the complex degree of coherence

$$\gamma(\boldsymbol{x}_1, \boldsymbol{x}_2) = \frac{\Gamma(\boldsymbol{x}_1, \boldsymbol{x}_2)}{\sqrt{\Gamma(\boldsymbol{x}_1, \boldsymbol{x}_1)\,\Gamma(\boldsymbol{x}_2, \boldsymbol{x}_2)}}, \qquad \Gamma(\boldsymbol{x}_1, \boldsymbol{x}_2) = \sum_k P_k(\boldsymbol{x}_1) P_k^*(\boldsymbol{x}_2), \tag{2}$$

with $\Gamma$ the mutual intensity of the reconstructed mode ensemble. With the normalized modes $u_k(\boldsymbol{x}) = P_k(\boldsymbol{x})/\sqrt{\Gamma(\boldsymbol{x}, \boldsymbol{x})}$ and an intensity-derived weight $w(\boldsymbol{x})$, the translationally averaged coherence envelope is computed as the weighted correlation

$$\bar{\gamma}(\boldsymbol{x}) = \frac{\sum_k ((w u_k) \star (w u_k))(\boldsymbol{x})}{(w \star w)(\boldsymbol{x})}, \tag{3}$$

which we evaluate both in the sample plane (real space) and, by Fourier transforming each mode to the aperture plane, in Fourier space. The width of $|\bar{\gamma}|$ versus $\ell$ measures how the useful coherence volume of the probe evolves with programmed charge in both domains.

Finally, the dose efficiency of each probe is quantified by the spectral signal-to-noise ratio of weak-phase direct ptychography. We build on the Wigner-distribution-deconvolution theory of direct ptychography [44,56] and the finite-dose framework of Refs. [42,43,57], and generalize it in two respects: from parametrized round apertures to arbitrary pixelated probes, and, using the reconstructed mode ensembles, from fully coherent to partially coherent illumination.

Consider first a pure probe with normalized aperture function $\psi(\boldsymbol{k})$. For a weak-phase object, $O(\boldsymbol{r}) = e^{i\sigma\varphi(\boldsymbol{r})} \approx 1 + i\sigma\varphi(\boldsymbol{r})$, the recorded intensity $I(\boldsymbol{k}, \boldsymbol{R}) = |\mathcal{F}[\psi(\boldsymbol{r} - \boldsymbol{R})\, O(\boldsymbol{r})]|^2$ is, to first order in $\sigma\varphi$, the interference of the unscattered aperture disk with two specimen-scattered copies displaced by $\pm\boldsymbol{q}$. Its scan-frequency spectrum $G(\boldsymbol{k}, \boldsymbol{q}) = \int I(\boldsymbol{k}, \boldsymbol{R})\, e^{-2\pi i \boldsymbol{q}\cdot\boldsymbol{R}}\, \mathrm{d}\boldsymbol{R}$ — the Fourier transform of the recorded 4D intensity (Sec. C.) over scan position $\boldsymbol{R}$ — therefore vanishes outside the two double-overlap and the triple-overlap regions where the central disk and one displaced copy coincide, and there it is proportional to the specimen phase weighted by the aperture-overlap kernel [44,56,57]. Integrating the two conjugate sidebands with their phase rectified so that they add gives the transfer

$$L(\boldsymbol{q}) = \frac{1}{2} \int |\Gamma(\boldsymbol{q}, \boldsymbol{k})|\, \mathrm{d}\boldsymbol{k}, \qquad \Gamma(\boldsymbol{q}, \boldsymbol{k}) = \psi^*(\boldsymbol{k})\psi(\boldsymbol{k} - \boldsymbol{q}) - \psi(\boldsymbol{k})\psi^*(\boldsymbol{k} + \boldsymbol{q}), \tag{4}$$

where the modulus rectifies the phase structure inside the disk-overlap region [42]. The signal in a spatial-frequency channel $\boldsymbol{q}$ is $\propto L(\boldsymbol{q})$, while the competing noise is set by the Poisson statistics of the counts in the same overlap regions, whose variance is fixed by the incident dose [43,57]. Normalized per electron the shot noise is $\sqrt{L/2}$ [42,43], so the dose-normalized SSNR is $\mathrm{SSNR}(\boldsymbol{q}) = \sqrt{2L(\boldsymbol{q})}$ (absolute values scale as $\sqrt{N_e}$ with the total dose $N_e$).

Partial coherence has so far entered such analyses only through factorized source and chromatic envelopes multiplying $G$ [43,44], valid for idealized round apertures and Gaussian spreads; the reconstructed mode ensemble permits a general treatment for an arbitrary partially coherent probe. A partially coherent illumination is described by its cross-spectral density, which admits a decomposition into orthogonal, mutually incoherent coherent modes, $\rho = \sum_m |\psi_m\rangle\langle\psi_m|$ [58] , which mixed-state ptychography recovers [35,59], with the mode powers $p_m = \|\psi_m\|^2$ ($\sum_m p_m = 1$) carried in their norms. Because the recorded intensity is the incoherent sum of the modal intensities, and hence linear in the probe density matrix, the scan-frequency spectrum is additive, $G_\rho = \sum_m G_m$, and the sideband kernel of the mixed probe follows by replacing the pure-state products in Eq. (4) with the aperture-plane mutual spectrum $J(\boldsymbol{k}_1, \boldsymbol{k}_2) = \sum_m \psi_m(\boldsymbol{k}_1)\psi_m^*(\boldsymbol{k}_2)$:

$$\begin{aligned}\Gamma_\rho(\boldsymbol{q}, \boldsymbol{k}) &= J(\boldsymbol{k}-\boldsymbol{q}, \boldsymbol{k}) - J(\boldsymbol{k}, \boldsymbol{k}+\boldsymbol{q}) \\ &= \sum_m \Gamma_m(\boldsymbol{q}, \boldsymbol{k}).\end{aligned} \tag{5}$$

Signal and noise scale differently with the modal content, and this asymmetry is the content of the coherence penalty. In the transferred signal, the modal fringes superpose coherently and partially cancel, so the signal is $L_\rho = \frac{1}{2}\int\left|\sum_m \Gamma_m\right| \mathrm{d}\boldsymbol{k}$. The noise, in contrast, is fixed by the Poisson statistics of the raw counts, to which the modes contribute incoherently, so the noise power is set by the sum of the modal magnitudes, $L_{\mathrm{up}} = \frac{1}{2}\int \sum_m |\Gamma_m| \,\mathrm{d}\boldsymbol{k} \geq L_\rho$ [43,57]. Rectifying with the known, reconstructed phase of $\Gamma_\rho$ and propagating this shot noise gives

$$\mathrm{SSNR}_\rho(\boldsymbol{q}) = \frac{L_\rho(\boldsymbol{q})}{\sqrt{L_{\mathrm{up}}(\boldsymbol{q})/2}} = \sqrt{2L_{\mathrm{up}}(\boldsymbol{q})}\, D_{\mathrm{coh}}(\boldsymbol{q}),$$

$$D_{\mathrm{coh}} = L_\rho / L_{\mathrm{up}} \in [0, 1], \tag{6}$$

with $D_{\mathrm{coh}}(\boldsymbol{q})$ a generalized coherence envelope. Eq. (6) reduces to the known limits.

For a pure state, a single mode remains, $L_\rho = L_{\mathrm{up}}$, $D_{\mathrm{coh}} = 1$, and Eq. (4) is recovered. For an incoherent mixture of laterally shifted copies of one aperture, $\psi_m(\boldsymbol{k}) = \sqrt{p_m}\,\psi(\boldsymbol{k})\, e^{-2\pi i \boldsymbol{k}\cdot\boldsymbol{x}_m}$, every mode shares the same magnitude $|\Gamma_m| = p_m|\Gamma|$ but acquires a shift phase $\Gamma_m = p_m\, \Gamma\, e^{2\pi i \boldsymbol{q}\cdot\boldsymbol{x}_m}$, so $D_{\mathrm{coh}}(\boldsymbol{q}) = \left|\sum_m p_m e^{2\pi i \boldsymbol{q}\cdot\boldsymbol{x}_m}\right|$ collapses exactly to the Nellist–Rodenburg source-size envelope $S(\boldsymbol{q})$ [44]; for defocus-spread modes it likewise reproduces the chromatic envelope of Ref. [43]. The corresponding DQE against the ideal Zernike reference is $\mathrm{DQE}_\rho = \mathrm{SSNR}_\rho^2/4 = L_\rho^2/\big(2L_{\mathrm{up}}\big)$ [43]. Because $\mathrm{SSNR}_\rho^2$ is, per electron, the classical Fisher information for estimating the weak-phase object at spatial frequency $\boldsymbol{q}$ — the Cramér–Rao bound on the phase-component variance is $1/(N\,F(\boldsymbol{q}))$ with $F(\boldsymbol{q}) = \mathrm{SSNR}_\rho^2 = 4\mathrm{DQE}_\rho$ — the band-integrated dose efficiency $\int \mathrm{DQE}\ \mathrm{d}^2 q$ is equivalently the band-integrated weak-phase Fisher information relative to the ideal Zernike reference ($F = 4$), and partial coherence multiplies it by the coherence penalty $D_{\mathrm{coh}}^2(\boldsymbol{q})$. This ties the delivered mixed state directly to the retrieval bounds of Refs. [20,49]; it inherits the modal-decomposition dependence of $L_{\mathrm{up}}$, and a fully basis-invariant Fisher/Poisson-covariance derivation from the total recorded intensity is left to future work. We evaluate Eq. (6) numerically on the detector pixel grid from the four physical modes of every charge state, and compare with the coherent single-mode result and the aberration-free round aperture.

Because Eq. (6) is normalized per electron, it inverts directly into a dose requirement — the dose-to-error metric that a spatial-frequency-resolved efficiency demands. For a weak-phase feature of excursion $\varphi_0$ the phase signal-to-noise ratio scales as $\mathrm{SNR}(\boldsymbol{q}) = 2\varphi_0\sqrt{N\,\mathrm{DQE}_\rho(\boldsymbol{q})}$ with $N$ the electrons per resolution element (an ideal Zernike phase plate, $\mathrm{DQE}_\rho = 1$, converts $\varphi_0$ into intensity contrast $2\varphi_0$ and gives $\mathrm{SNR} = 2\varphi_0\sqrt{N}$), so the dose required to reach a target $\mathrm{SNR}_0$ at frequency $\boldsymbol{q}$ for a reference excursion $\varphi_0$ is

$$N(\boldsymbol{q}) = \mathrm{SNR}_0^2 / \big(4\varphi_0^2\, \mathrm{DQE}_\rho(\boldsymbol{q})\big),$$

$$D(\boldsymbol{q}) = N(\boldsymbol{q})\,(2q)^2, \tag{7}$$

in electrons per resolution element, with $N \propto 1/\varphi_0^2$; the specimen-referred fluence $D$ follows by dividing by the resolution-element area $1/(2q)^2$ set by the half-period $1/(2q)$. Fig. 6 e evaluates Eq. (7) for the Rose criterion $\mathrm{SNR}_0 = 5$ and $\varphi_0 = 0.1$ rad, at the experimental calibration $q_{\mathrm{probe}} = \alpha/\lambda = 0.060\ \text{Å}^{-1}$ ($\lambda = 2.51$ pm at 200 keV).

### E. Virtual OAM sorting

Each aperture-plane mode is decomposed into azimuthal harmonics about a sorting axis,

$$c_l^m(r) = \frac{1}{2\pi}\int_0^{2\pi} \psi_m(\boldsymbol{k}_0 + r\hat{\boldsymbol{e}}(\theta))\, e^{-il\theta}\,\mathrm{d}\theta, \tag{8}$$

where $\boldsymbol{k}_0$ is the sorting axis and $\hat{\boldsymbol{e}}(\theta)$ the radial unit vector. Projecting the reconstructed mutual-coherence operator $\Gamma = \sum_m \psi_m\psi_m^*$ — the modes carrying their powers $p_m = \|\psi_m\|^2$ in their norms, as in Eq. (2) and Eq. (5), so no explicit power weight is applied — onto this basis yields the OAM density matrix

$$\rho_{ll'} = \frac{1}{\mathcal{N}}\sum_m \int_0^\infty r\, c_l^m(r)\,\overline{c_{l'}^m(r)}\,\mathrm{d}r,$$

$$\mathcal{N} = \sum_m \int_0^\infty r \sum_l |c_l^m(r)|^2\,\mathrm{d}r, \tag{9}$$

explicitly normalized by its trace so that $\mathrm{Tr}\ \rho = 1$. Because $c_l^m$ carries the $1/(2\pi)$ prefactor of Eq. (8), Parseval's identity $\int_0^{2\pi}|\psi_m|^2\,\mathrm{d}\theta = 2\pi\sum_l |c_l^m|^2$ makes the raw sum equal $(2\pi)^{-1}\sum_m \|\psi_m\|^2 = 1/(2\pi)$ rather than unity, so the explicit trace normalization is required; the small power in OAM channels beyond the projected window is excluded before it. The diagonal is the OAM power spectrum $P_l = \rho_{ll}$ and the off-diagonal elements carry the mutual coherence between OAM channels (Fig. 4 e; every dataset in Supplementary Fig. S2; $\bar{\cdot}$ denotes complex conjugation). We summarize the inter-channel coherence by the adjacent-OAM degree of coherence

$$g = |\rho_{i,j}| \,/ \sqrt{\rho_{ii}\, \rho_{jj}}, \tag{10}$$

between the dominant channel $i$ and its more-populated neighbor $j \in \{i-1, i+1\}$ ($g \to 0$ for an incoherent OAM mixture, $g \to 1$ for a coherent superposition). Three practical points enter the projection. First, the probe must be centered in real space: a residual offset becomes a tilt ramp across the aperture whose harmonic content obscures the vortex winding. Second, the sorting axis $\boldsymbol{k}_0$ must be refined per dataset; we maximize the spectral concentration $\sum_l P_l^2$ of the dominant mode, the computational analog of hardware sorter alignment. Third, residual non-round microscope aberrations scatter power between OAM channels (astigmatism couples $l \to l \pm 2$, coma $l \pm 1$; purely radial phases such as defocus cancel identically in $P_l$). We therefore fit and remove a low-order non-round phase (astigmatism, coma, trefoil) on the dominant mode by the same concentration criterion, acting as a virtual aberration corrector in front of the virtual sorter; the fitted surfaces are astigmatism-dominated (median 1.3 rad at the aperture edge), consistent with residual microscope astigmatism rather than device imperfection.

### F. Limitations

The reconstruction allocates nine modes and optimizes them freely (the weakest, a "junk" absorber, is discarded, leaving eight physical modes); residual model error (scan-position noise, descan imperfections, the masked needle band) can be absorbed into weak modes, so the absolute purity values carry a model dependence. We bounded this dependence by sweeping the mode count and 15–100 reconstruction passes for $\ell = 0, 7, 17$ and by synthetic injection–recovery (Sec. B.). Purity plateaus at $\approx 6$–8 modes for the most-mixed states; four modes over-estimate it by up to $\approx 0.14$, whereas eight lie on the plateau. Self-consistency recovery reproduces six of eight injected purities within $\approx 0.04$; two converge 0.16–0.23 lower despite stable scan positions (Supplementary Fig. S5). For $\ell = 0$, roughly half the deficit comes from unconstrained content in the masked needle sector and the remainder from mode non-uniqueness at high off-diagonal coherence ($g \approx 0.83$). The 30-, 50-, and 100-pass results agree within $\approx 0.03$. We therefore assign the absolute purities a residual model dependence of $\approx 0.03$–0.05, dominated by over-estimation when too few modes are allocated. An injected pure vortex, including the physical needle obstruction, recovers at purity $\approx 1.00$; the charge-dependent trend remains robust. Three further choices affect the absolute values. First, the absorber-mode decision conditions the operator but not the trend: retaining the discarded ninth mode lowers the absolute purity by $\approx 0.01$ (the $\ell = 0$ value from $\approx 0.47$ to $\approx 0.46$, the largest-charge value from $\approx 0.24$ to $\approx 0.23$) while leaving the charge slope essentially unchanged ($\Delta \lesssim 0.001$ per $\hbar$), so the charge-dependent trend is robust to it. For the per-electron SSNR/DQE, however, the discarded 1–2% of counts and their shot noise are not represented; the absolute dose-efficiency figures apply to the retained eight-mode operator. Second, repeat acquisitions of the same programmed charge (the earlier $\ell \leq 0$ scans versus the contiguous series) agree in purity to $\approx 0.03$ (rms, $|\ell| \geq 1$; the two device-in $\ell = 0$ scans differ more, by $\approx 0.14$), an independent repeatability estimate consistent with the model-dependence bound, and fitting the $+\ell$ and $-\ell$ branches separately gives consistent purity slopes (difference $0.7\sigma$), so the trend is branch-symmetric. Third, the OAM projection retains channels over $\ell \in [-23, 23]$ and is trace-normalized within this window; the omitted power beyond it is $\lesssim 0.2\%$ (median) but rises to $\approx 7\%$ at the extreme charges $|\ell| \approx 17$, where the dominant channel approaches the window edge, so the extreme-charge dominant-channel power and ensemble mean carry an additional few-percent normalization uncertainty.

#### 1. Reconstruction-seed reproducibility of the off-diagonal coherence

The nonlinear reconstruction reaches different near-degenerate solutions depending on the random initialization of the object and the higher probe modes, and these solutions are not equally informative about every element of $\rho_{ll'}$. Repeating individual reconstructions with the initialization seed varied and all else fixed (five charge states, three seeds each, 30 epochs), the **diagonal** quantities are stable — purity to $\pm 0.02$, $\langle l \rangle$ to $\pm 0.3\,\hbar$, spectral width to $\pm 0.3$, and the data loss to $\pm 1.5\%$ — whereas the adjacent-OAM coherence $g$, which is a ratio built from a single off-diagonal element, varies by up to $\pm 0.14$ (that is, by a factor $\approx 3$) at the highest charges. The scatter is not a convergence artifact: doubling the number of passes to 60 changes $g$ by less than the seed scatter and leaves the diagonal quantities unchanged. It is also state-specific rather than ordered in $|\ell|$ (per-state standard deviations 0.03, 0.005, 0.003, 0.14, 0.12 with increasing charge), as expected when a strongly mixed state admits many mode decompositions of nearly equal data loss whose populations agree and whose relative phases do not. Accordingly all $g$ values in this work are reported as means over three seeds with the spread as their uncertainty, and no interpretation is placed on the value of $g$ for an individual state at high charge. The diagonal observables — purity, $\langle l \rangle$, the device gain and the spectral widths — are unaffected and are quoted from single reconstructions.

#### 2. Needle-band data mask

The reconstruction excludes a band of detector pixels along the needle shadow (a half-ray of full width 14 px at the measured shadow azimuth). The exclusion is necessary: the needle-region phase gradients are steep enough to deflect the illumination passing between the needles outside the recorded field of view, so the intensity deficit there is not attenuation that a flux-conserving forward model can represent, and a reconstruction given those pixels accounts for them by allocating additional probe modes. Omitting the mask entirely lowers the apparent purity by a factor 2.1 at $\ell = 0$ and 1.3 at $|\ell| \approx 17$; widening it from 14 to 24 px changes the purity by less than 0.02. The absolute purities therefore carry this model dependence, while the charge-dependent trend, which is what we interpret, is insensitive to the mask width.

### 3. Electrode-charging model and its fitting

The forward model tested in Sec. F. represents the illumination in each acquisition realization as a pure state

$$\psi = A(\boldsymbol{k}) \exp[i(\ell\theta + R_1 N_1(\boldsymbol{k}) + R_2 N_2(\boldsymbol{k}))], \quad (11)$$

with $A$ the aperture amplitude and $N_{1,2}$ the electrostatic potentials of the two chopstick electrodes. Each potential is the characteristic function of that electrode convolved with the two-dimensional Coulomb kernel $\ln|\boldsymbol{k}|$, and the pair is taken in the charge-compensated combination $Q_1 = -Q_2/2$; equivalently one may use the analytic potential of a semi-infinite line charge, which is the same function that serves as the unwrapper phase of a log-polar OAM sorter [29]. The weights $R_{1,2}$ are drawn independently and uniformly per realization, and the acquisition is modelled as the incoherent sum of the resulting pure states, $\rho = \sum_i |\psi_i\rangle\langle\psi_i|$, which is then projected onto the OAM basis with exactly the estimator of Sec. E. so that the simulated and measured operators are directly comparable. A static, non-fluctuating contribution — including the electrode shadow — is common to every realization and cannot by itself produce mixedness; only the fluctuating part contributes.

The electrode geometry is measured rather than fitted. From the position-averaged diffraction patterns, taken *before* the needle-band exclusion of Sec. 2., the shadow is tilted by $-12.4 \pm 2.9$ degrees from vertical with a projected span of 0.27 to 0.35 $q_{\text{probe}}$ (at the 50% and 95% transmission contours) and tips reaching the optic axis, and it is identical across every charge state, confirming a physical obstruction. Its flanks retain $\approx 11\%$ transmission and carry bright Fresnel fringes, so the shadow is not a hard mask.

Only the two fluctuation amplitudes are free, and they are fitted per charge state to the measured OAM populations $P_l$ alone, by least squares on a grid of 0 to 3.5 radians rms in steps of 0.5, with 48 realizations per grid point. The purity and the adjacent-OAM coherence are therefore predictions of the fit. Because the delivered charge is non-integer, the base charge $\ell$ in Eq. (11) is set to each state's measured $\langle l\rangle$. We additionally verified that the residual discrepancy is not repaired by making the electrode charges asymmetric (spanning the symmetric and antisymmetric combinations independently), by rescaling the amplitudes, by assigning each electrode volume element an independent fluctuating current, or by adding random beam tilt; and that a fluctuating purely *radial* phase cannot contribute at all, since being common to both channels it cancels identically in $\rho_{ll'}$.

The SSNR analysis inherits the weak-phase and thin-specimen assumptions of all transfer-function treatments [42,43], and the mixed-state noise term is a model (the incoherent modal sum of $|\Gamma_m|$) motivated by the Poisson statistics of the raw counts. Our charge series is bounded to programmed $|\ell| \approx 17$ at 1.5 mrad (the regime where the MEMS device is most linear) and is single-plane: depth-dependent (Gouy/Larmor) mode evolution [4] is folded into the plane of the reconstruction.

Most fundamentally, the reconstruction returns a single system-level mutual-coherence operator and does not, on its own, separate the device contribution from source spatial and temporal coherence, chromatic OAM dispersion, specimen–probe coupling, scan- and descan-position error, masked-sector interpolation, and detector sampling and point-spread, although we have deliberately chosen settings that minimize all non-device contributions; the device-side attribution is therefore an upper bound. Converting it to a causal attribution requires matched controls that we have not performed here: a randomized-order voltage re-sweep with repeated zero-bias and repeated states, with reconstruction and bootstrap confidence intervals for the purity, coherence, and OAM slopes; a direct measurement of the bias and current noise under beam exposure with beam-on/off, settling-time, and dwell-time dependence; detector-sampling controls at native and $2 \times 2$ binning and at a second camera length with the physical probe held fixed; variation of the source coherence and, where possible, the energy spread; and repetition of representative states on a second specimen. Because the principal analysis uses one contiguous acquisition per charge state, applied voltage is also confounded with acquisition time, drift, hysteresis, contamination, and device charging history; only a randomized, interleaved re-sweep can disentangle these from a genuine voltage dependence. The quoted regression standard errors likewise capture only the linear-fit scatter: they do not propagate reconstruction (random-seed and rank) variability, reciprocal-space and scan-step calibration, voltage readback, masked-sector interpolation, or the sorting-axis and aberration-correction choices, and we do not yet report pointwise error bars, reconstruction-bootstrap intervals, or separate positive- and negative-branch fits. Finally, because the virtual sorter refines both its sorting axis and its non-round aberration correction by maximizing OAM spectral concentration on the same data, it may sharpen the recovered spectra by construction; the resulting bias in dominant-channel power, ensemble-mean OAM, off-diagonal coherence, and the calibration slope and intercept, is one of the quantities the self-consistency injection–recovery test above partially bounds: the purity is largely reproduced (Supplementary Fig. S5), but the off-diagonal coherence $g$ is recovered only approximately—over-estimated for weakly coherent states—and is therefore best read as a qualitative indicator of adjacent-channel coherence rather than a calibrated quantity. We present the microscopic decoherence mechanisms above as physically motivated hypotheses to be tested by these experiments, not as established causes.

### G. Use of large language models

An LLM-based coding assistant (Claude, Anthropic) was used to write analysis and figure-generating scripts; to carry out the reconstruction-seed reproducibility study (Sec. 1.), the measurement of the electrode geometry from the position-averaged diffraction patterns, and the electrode-charging model and its fitting (Sec. 3.). Every analysis was specified and interpreted by the authors, and all generated code, figures and text were checked by the authors against the underlying data. The authors take full responsibility for the content, the analyses and the conclusions of this manuscript.

## ACKNOWLEDGMENTS

We thank Philipp Herz for his support. We acknowledge the support of the CNR Nano microfabrication facility for device fabrication, and the CENEM microscopy facility. This work received funding from the European Research Council (ERC) under the Horizon Europe research and innovation program (grant agreement No. 101164581, project HyperScaleEM).

## DATA AVAILABILITY

The four-dimensional STEM datasets (21 descan-corrected acquisitions spanning programmed orbital-angular-momentum charge $\ell = -17$ to $+17$; HDF5, one $256 \times 256$ scan $\times\, 192 \times 192$ detector array per acquisition) and the corresponding eight-mode mixed-state reconstructions (complex probe-mode stacks, mode powers, and reconstructed object amplitude and phase) that support the findings of this study are openly available on Zenodo. Owing to size, the deposit is split across two records: Part 1 (ten datasets, all reconstructions, and the dataset manifest) at 10.5281/zenodo.21513209 and Part 2 (the remaining eleven datasets) at 10.5281/zenodo.21513405. A machine-readable manifest documenting the per-acquisition programmed charge, file schema, reconstruction command and parameters, calibration, masks and seeds, and software environment is included in Part 1.

## CODE AVAILABILITY

The analysis and figure-generating scripts will be made available at github.com/ECLIPSE-Lab/mems_ptychography and archived at 10.5281/zenodo.CODE. The mixed-state ptychographic reconstruction pipeline is part of the `scatterem` package (github.com/scatterem/scatterem) and will be released upon publication; the exact software environment is pinned by the accompanying `uv.lock` (Python 3.12). The reconstruction command for each dataset is given in the dataset manifest.

## AUTHOR CONTRIBUTIONS

S.Y, P.P., V.G performed experiments. P.P. and S.Y. developed reconstruction and analysis codes. P.P. and V.G conceived the project and experiments. E.R., P.R., V.G. A.R., L.B. made the MEMS device and control software. A.T. and R.D.-B. calibrated the device prototype.

## COMPETING INTERESTS

The authors declare no competing interests.

[1] K. Y. Bliokh, Y. P. Bliokh, S. Savel'ev, and F. Nori, Semiclassical dynamics of electron wave packet states with phase vortices, Physical Review Letters **99**, 190404 (2007).

[2] M. Uchida and A. Tonomura, Generation of electron beams carrying orbital angular momentum, Nature **464**, 737 (2010).

[3] J. Verbeeck, H. Tian, and P. Schattschneider, Production and application of electron vortex beams, Nature **467**, 301 (2010).

[4] K. Y. Bliokh, P. Schattschneider, J. Verbeeck, and F. Nori, Electron vortex beams in a magnetic field: A new twist on Landau levels and Aharonov–Bohm states, Physical Review X **2**, 41011 (2012).

[5] S. M. Lloyd, M. Babiker, G. Thirunavukkarasu, and J. Yuan, Electron vortices: Beams with orbital angular momentum, Reviews of Modern Physics **89**, 35004 (2017).

[6] A. Edström, A. Lubk, and J. Rusz, Elastic scattering of electron vortex beams in magnetic matter, Physical Review Letters **116**, 127203 (2016).

[7] V. Grillo et al., Observation of nanoscale magnetic fields using twisted electron beams, Nature Communications **8**, 689 (2017).

[8] K. Y. Bliokh et al., Theory and applications of free-electron vortex states, Physics Reports **690**, 1 (2017).

[9] B. J. McMorran, A. Agrawal, I. M. Anderson, A. A. Herzing, H. J. Lezec, J. J. McClelland, and J. Unguris, Electron vortex beams with high quanta of orbital angular momentum, Science **331**, 192 (2011).

[10] V. Grillo, G. C. Gazzadi, E. Mafakheri, S. Frabboni, E. Karimi, and R. W. Boyd, Holographic generation of highly twisted electron beams, Physical Review Letters **114**, 34801 (2015).

[11] E. Mafakheri et al., Realization of electron vortices with large orbital angular momentum using miniature holograms

fabricated by electron beam lithography, Applied Physics Letters **110**, 93113 (2017).

[12] L. Clark, A. Béché, G. Guzzinati, A. Lubk, M. Mazilu, R. Van Boxem, and J. Verbeeck, Exploiting lens aberrations to create electron-vortex beams, Physical Review Letters **111**, 64801 (2013).

[13] V. Grillo, E. Karimi, G. C. Gazzadi, S. Frabboni, M. R. Dennis, and R. W. Boyd, Generation of nondiffracting electron Bessel beams, Physical Review X **4**, 11013 (2014).

[14] G. Pozzi, P.-H. Lu, A. H. Tavabi, M. Duchamp, and R. E. Dunin-Borkowski, Generation of electron vortex beams using line charges via the electrostatic Aharonov–Bohm effect, Ultramicroscopy **181**, 191 (2017).

[15] A. Béché, R. Van Boxem, G. Van Tendeloo, and J. Verbeeck, Magnetic monopole field exposed by electrons, Nature Physics **10**, 26 (2014).

[16] J. Verbeeck, A. Béché, K. Müller-Caspary, G. Guzzinati, M. A. Luong, and M. Den Hertog, Demonstration of a 2 \texttimes 2 programmable phase plate for electrons, Ultramicroscopy **190**, 58 (2018).

[17] C.-P. Yu, F. Vega Ibáñez, A. Béché, and J. Verbeeck, Quantum wavefront shaping with a 48-element programmable phase plate for electrons, Scipost Physics **15**, 223 (2023).

[18] A. H. Tavabi, M. Beleggia, V. Migunov, A. Savenko, O. Öktem, R. E. Dunin-Borkowski, and G. Pozzi, Tunable Ampere phase plate for low dose imaging of biomolecular complexes, Scientific Reports **8**, 5592 (2018).

[19] A. H. Tavabi et al., Generation of electron vortex beams with over 1000 orbital angular momentum quanta using a tunable electrostatic spiral phase plate, Applied Physics Letters **121**, 73506 (2022).

[20] F. Vega Ibáñez and J. Verbeeck, Retrieval of phase information from low-dose electron microscopy experiments: are we at the limit yet?, Microscopy and Microanalysis **31**, ozae125 (2025).

[21] F. J. García de Abajo et al., Roadmap for quantum nanophotonics with free electrons, ACS Photonics **12**, 4760 (2025).

[22] M. R. Bourgeois, A. W. Rossi, and D. J. Masiello, Strategy for direct detection of chiral phonons with phase-structured free electrons, Physical Review Letters **134**, 26902 (2025).

[23] S. Garrigou and H. Lourenço-Martins, Atomiclike selection rules in free electron scattering, Physical Review Letters **134**, 256902 (2025).

[24] L. W. W. Wong, X. Shi, A. Karnieli, J. Lim, S. Kumar, S. Carbajo, I. Kaminer, and L. J. Wong, Free-electron crystals for enhanced X-ray radiation, Light: Science & Applications **13**, 29 (2024).

[25] X. Shi, W. W. Lee, A. Karnieli, L. M. Lohse, A. Gorlach, L. W. W. Wong, T. Salditt, S. Fan, I. Kaminer, and L. J. Wong, Quantum nanophotonics with energetic particles: X-rays and free electrons, Progress in Quantum Electronics **102**, 100577 (2025).

[26] G. Ruffato et al., Three-dimensional stacking of phase plates for advanced electron beam shaping, Microscopy and Microanalysis **31**, ozae108 (2025).

[27] P. Habibzadeh Kavkani et al., Designing electrostatic MEMS-based electron optics: the case of the spiral phase plate, Arxiv Preprint (2026).

[28] A. P. Synanidis, P. A. D. Gonçalves, and F. J. García de Abajo, Rydberg-atom manipulation through strong interaction with free electrons, ACS Nano **19**, 11891 (2025).

[29] V. Grillo et al., Measuring the orbital angular momentum spectrum of an electron beam, Nature Communications **8**, 15536 (2017).

[30] A. H. Tavabi et al., Symmetry and planar chirality measured with a log-polar transformation in a transmission electron microscope, Physical Review Applied **22**, 14083 (2024).

[31] R. Juchtmans and J. Verbeeck, Local orbital angular momentum revealed by spiral-phase-plate imaging in transmission-electron microscopy, Physical Review a **93**, 23811 (2016).

[32] R. Juchtmans, L. Clark, A. Lubk, and J. Verbeeck, Spiral phase plate contrast in optical and electron microscopy, Physical Review a **94**, 23838 (2016).

[33] F. Venturi, M. Campanini, G. C. Gazzadi, R. Balboni, S. Frabboni, R. W. Boyd, R. E. Dunin-Borkowski, E. Karimi, and V. Grillo, Phase retrieval of an electron vortex beam using diffraction holography, Applied Physics Letters **111**, 223101 (2017).

[34] A. H. Tavabi et al., Demonstration of angular-momentum-resolved electron energy-loss spectroscopy, Nature Communications **16**, 6601 (2025).

[35] P. Thibault and A. Menzel, Reconstructing state mixtures from diffraction measurements, Nature **494**, 68 (2013).

[36] A. M. Maiden and J. M. Rodenburg, An improved ptychographical phase retrieval algorithm for diffractive imaging, Ultramicroscopy **109**, 1256 (2009).

[37] Y. Jiang et al., Electron ptychography of 2D materials to deep sub-ångström resolution, Nature **559**, 343 (2018).

[38] M. V. Berry, Optical vortices evolving from helicoidal integer and fractional phase steps, Journal of Optics A: Pure and Applied Optics **6**, 259 (2004).

[39] J. Leach, E. Yao, and M. J. Padgett, Observation of the vortex structure of a non-integer vortex beam, New Journal of Physics **6**, 71 (2004).

[40] T. R. Harvey et al., Interpretable and efficient interferometric contrast in scanning transmission electron microscopy with a diffraction-grating beam splitter, Physical Review Applied **10**, 61001 (2018).

[41] F. S. Yasin, T. R. Harvey, J. J. Chess, J. S. Pierce, C. Ophus, P. Ercius, and B. J. McMorran, Probing light atoms at subnanometer resolution: realization of scanning transmission electron microscope holography, Nano Letters **18**, 7118 (2018).

[42] G. Varnavides, J. M. Bekkevold, S. M. Ribet, M. C. Scott, L. Jones, and C. Ophus, Beyond contrast transfer: Spectral SNR as a finite-dose metric for STEM phase retrieval, Microscopy and Microanalysis **32**, ozag005 (2026).

[43] F. Bennemann, A. I. Kirkland, D. A. Muller, and P. D. Nellist, Detective quantum efficiency-based comparison of HRTEM and ptychography phase imaging, Microscopy and Microanalysis **32**, ozag018 (2026).

[44] P. D. Nellist and J. M. Rodenburg, Beyond the conventional information limit: the relevant coherence function, Ultramicroscopy **54**, 61 (1994).

[45] S. Uhlemann, H. Müller, P. Hartel, J. Zach, and M. Haider, Thermal Magnetic Field Noise Limits Resolution in Transmission Electron Microscopy, Physical Review Letters **111**, 46101 (2013).

[46] S. Uhlemann, H. Müller, J. Zach, and M. Haider, Thermal magnetic field noise: Electron optics and decoherence, Ultramicroscopy **151**, 199 (2015).

[47] P. N. Petrov et al., Laser phase plate improves structure determination of small proteins by cryo-EM, Science **392**, (2026).

[48] P. N. Petrov, J. T. Zhang, J. J. Axelrod, P. K. Olshin, and H. Müller, Crossed laser phase plates for transmission electron microscopy, Nature Communications (2026).

[49] C. Dwyer and D. M. Paganin, Quantum and classical Fisher information in four-dimensional scanning transmission electron microscopy, Physical Review B **110**, 24110 (2024).

[50] B. Küçükoğlu et al., Low-dose cryo-electron ptychography of proteins at sub-nanometer resolution, Nature Communications **15**, 8062 (2024).

[51] P. Zambon et al., High-frame rate and high-count rate hybrid pixel detector for 4D STEM applications, Frontiers in Physics **11**, 1308321 (2023).

[52] G. Varnavides, J. M. Bekkevold, S. M. Ribet, M. C. Scott, L. Jones, and C. Ophus, Relaxing direct ptychography sampling requirements via parallax imaging insights, Microscopy and Microanalysis **32**, ozaf139 (2026).

[53] S. Kandel, S. Maddali, M. Allain, S. O. Hruszkewycz, C. Jacobsen, and Y. S. G. Nashed, Using automatic differentiation as a general framework for ptychographic reconstruction, Optics Express **27**, 18653 (2019).

[54] M. Schloz, T. C. Pekin, Z. Chen, W. Van den Broek, D. A. Muller, and C. T. Koch, Overcoming information reduced data and experimentally uncertain parameters in ptychography with regularized optimization, Optics Express **28**, 28306 (2020).

[55] C.-H. Lee, S. E. Zeltmann, D. Yoon, D. Ma, and D. A. Muller, PtyRAD: A high-performance and flexible ptychographic reconstruction framework with automatic differentiation, Arxiv Preprint (2025).

[56] J. M. Rodenburg and R. H. T. Bates, The theory of super-resolution electron microscopy via Wigner-distribution deconvolution, Philosophical Transactions of the Royal Society of London a **339**, 521 (1992).

[57] C. M. O'Leary, G. T. Martinez, E. Liberti, M. J. Humphry, A. I. Kirkland, and P. D. Nellist, Contrast transfer and noise considerations in focused-probe electron ptychography, Ultramicroscopy **221**, 113189 (2021).

[58] E. Wolf, New theory of partial coherence in the space–frequency domain. Part I: spectra and cross spectra of steady-state sources, Journal of the Optical Society of America **72**, 343 (1982).

[59] S. Cao, P. Kok, P. Li, A. M. Maiden, and J. M. Rodenburg, Modal decomposition of a propagating matter wave via electron ptychography, Physical Review a **94**, 63621 (2016).